\documentclass[12pt,onecolumn,oneside,peerreview]{IEEEtran}

\usepackage{graphicx}
\usepackage{cite}
\usepackage{float}
\usepackage{amsmath}
\usepackage{times}
\usepackage{latexsym}
\usepackage{bm}
\usepackage{amssymb}
\usepackage[center]{caption2}
\usepackage{stfloats}
\usepackage{cases}
\usepackage{array}
\usepackage{setspace}
\usepackage{fancyhdr}
\usepackage{color}
\usepackage{subfigure}
\usepackage{multirow}
\usepackage{amsthm}
\usepackage{cases}
\usepackage{algorithmic,algorithm}
\usepackage[long]{optidef}
\usepackage{epstopdf}
\usepackage{verbatim}
\usepackage{amsmath}
\allowdisplaybreaks[4]

\usepackage{amsmath}
\newtheorem{remark}{Remark}

\renewcommand{\baselinestretch}{0.97}

\ifCLASSINFOpdf

\else

\fi

\begin{document}

\title{\color{black}Sensing-based Beamforming Design for Joint Performance Enhancement of RIS-Aided ISAC Systems}

\author{\IEEEauthorblockN{Xiaowei Qian, {\em Graduate Student Member, IEEE}, Xiaoling Hu, {\em Member, IEEE}, \\Chenxi Liu, {\em Senior Member, IEEE}, Mugen Peng, {\em Fellow, IEEE}\\ and Caijun Zhong, {\em Senior Member, IEEE}}

\thanks{Xiaowei Qian, Xiaoling Hu, Chenxi Liu and Mugen Peng are with the State Key Laboratory of Networking and Switching Technology, Beijing University of Posts and Telecommunications, Beijing 100876, China (e-mail: \{xiaowei.qian, xiaolinghu, chenxi.liu, pmg\}@bupt.edu.cn).}
\thanks{Caijun Zhong is with the College of Information Science and Electronic Engineering, Zhejiang University, Hangzhou 310027, China (email: caijunzhong@zju.edu.cn).}
}
\maketitle

\begin{abstract}
Reconfigurable intelligent surface (RIS) has shown its great potential in facilitating device-based integrated sensing and communication (ISAC), where sensing and communication tasks are mostly conducted on different time-frequency resources. While the more challenging scenarios of simultaneous sensing and communication (SSC) have so far drawn little attention. In this paper, we propose a novel RIS-aided ISAC framework where the inherent location information in the received communication signals from a blind-zone user equipment is exploited to enable SSC. We first design a two-phase ISAC transmission protocol. In the first phase, communication and coarse-grained location sensing are performed concurrently by exploiting the very limited channel state information, while in the second phase, by using the coarse-grained sensing information obtained from the first phase, simple-yet-efficient sensing-based beamforming designs are proposed to realize both higher-rate communication and fine-grained location sensing. We demonstrate that our proposed framework can achieve almost the same performance as the communication-only frameworks, while providing up to millimeter-level positioning accuracy. In addition, we show how the communication and sensing performance can be simultaneously boosted through our proposed sensing-based beamforming designs. The results presented in this work provide valuable insights into the design and implementation of other ISAC systems considering SSC.
\end{abstract}

\begin{IEEEkeywords}
  Integrated sensing and communication, reconfigurable intelligent surface, multiple-input-multiple-output system, location estimation, beamforming design.
\end{IEEEkeywords}

\vspace{-5mm} \section{Introduction}
Integrated sensing and communication (ISAC) has been regarded as one of the potential technologies for the sixth-generation (6G) networks\cite{liu2022integrated}, providing both high-accuracy sensing services and high-rate communication services to support visionary applications such as internet of things, autonomous vehicles, extended reality, etc.. Generally, sensing and communication systems are designed separately, and operate in different frequency bands. However, with the development of the millimeter-wave (mmWave) and massive multiple-in-multiple-out (MIMO) technologies, communication signals in mmWave feature high-resolution sensing in both time and angular domains, making it possible to realize high-accuracy sensing by using communication signals\cite{liu2022survey}. This motivates ISAC systems, which aims to perform sensing and communication tasks simultaneously in a single system such that both higher spectrum efficiency and lower hardware cost can be achieved\cite{liu2020joint}. \textcolor{black}{Moreover, the ISAC systems can be integrated at different levels, from loosely coupled to fully integrated, and from mutual competition to mutual enhancement\cite{tan2021integrated}}. In general, the ISAC can be divided into two categories, i.e., device-based ISAC and device-free ISAC, depending on whether the sensing objective is cooperative (e.g. the user in the mobile networks) or non-cooperative (e.g. birds or unmanned aerial vehicles)\cite{liu2022integrated}.

MmWave band is the ideal frequency band for ISAC systems, which can provide enormous bandwidth to enhance both the transmission rate and sensing accuracy in ISAC systems. However, the pathloss in the mmWave band is extremely severe, which weakens signals from non-line-of-sight (NLoS) paths and causes blind zones in the networks, harming both sensing and communication performances. To cope with these issues, some traditional solutions such as ultra-dense networks (UDN) \cite{zhang2017energy} and amplify-and-forward (AF) relaying \cite{jiang2019mmwave} were proposed. Whereas, these solutions inevitably require extra radio-frequency (RF) hardware, causing additional energy consumption and inducing unwanted interference. \textcolor{black}{Beamforming design based on artificial intelligence has also been proposed to coup with such issues \cite{R2-added-reviewer5}.} Recently, the emergence of the reconfigurable intelligent surface (RIS) provides a new way to tackle these challenges in mmWave ISAC systems \cite{liu2021reconfigurable}. The RIS, which can manipulate the wireless propagation environment, is essentially a programmable electromagnetic meta-surface with the structure of a two-dimensional planar array composed of a large amount of low-cost passive reflecting elements, which are able to manipulate the phase and the amplitude of the incident signals \cite{cui2014coding,liaskos2018new,gong2020toward}. Thereby, the RIS is capable of passively reflecting the incident signals to establish a strong virtual line-of-sight (LoS) link between the transmitter and the receiver without any RF hardware and additional energy consumption \cite{bjornson2019intelligent,pan2021reconfigurable}, which is attractive for mmWave ISAC systems.

Due to the above advantages of utilizing RISs in ISAC systems, RIS-aided ISAC systems have gained extensive research interests in the existing literature. In RIS-aided device-free ISAC systems, the RIS is mainly used to suppress the mutual interference between sensing and communication systems \cite{wang2020ris,he2022ris} or \textcolor{black}{multi-user interference (MUI) \cite{wei2022multiple,jiang2021intelligent,yan2022reconfigurable,wangxinyi2021joint,R2-added-2023-RIS-ISAC}}. For example, the work \cite{wang2020ris} addressed the spectrum sharing problem between MIMO radar and multi-user communication systems, where RIS is exploited to deal with the interference imposed by the base station (BS), while the work \cite{wangxinyi2021joint} studied the minimization of MUI under the strict beampattern constraint by jointly optimizing the ISAC waveform and the RIS phase shift matrix and demonstrated that the system throughput can be significantly improved with the RIS. In RIS-aided device-based ISAC systems, only a few works have investigated the joint communication and sensing (JC\&S) for a mobile user equipment (UE). However, such system cannot be ignored for its potential of easy-to-deploy and mutual synergy. For example, the work \cite{heJ2022beyond} introduced the fundamentals of RIS mmWave channel modeling, followed by RIS channel state information (CSI) acquisition. It concluded that the ISAC is beneficial to harness the full potential of RISs compared with separate sensing and communication systems. \textcolor{black}{The work \cite{wangrui2021joint} established an RIS-aided mmWave-MIMO based JC\&S system and then derived the approximate closed-form expressions of its location sensing error bound and achievable data rate. Then, a JC\&S metric was proposed followed by an optimization algorithm to find the optimal time allocation ratio that maximizes the JC\&S performance.} Furthermore, unlike the works \cite{heJ2022beyond,wangrui2021joint} that conduct communication and localization tasks on different time resources, the work \cite{liu2022sensing} proposed an RIS-aided multiple-in-single-out (MISO) ISAC framework, where communication and location sensing tasks for a blind-zone user are conducted on the same time-frequency resources, with the aid of a distributed semi-passive RIS. Location sensing and beamforming design schemes are respectively proposed. Utilizing the imperfect location information obtained in the proposed location sensing scheme, the proposed beamforming design scheme can achieve comparable communication performance to the optimal beamforming scheme with perfect CSI.

Most of the above works assume that the perfect CSI is available for RIS-aided ISAC systems. Whereas, the acquisition of the CSI is still a challenging issue for the RIS-aided systems due to the large number of coefficients in the cascaded BS-RIS-UE channel. Although, some channel estimation schemes are proposed to estimate the cascaded channel by schematically maneuvering all or part of the RIS reflecting elements, such as \cite{nadeem2019intelligent,zheng2019intelligent,chen2019channel}, the overhead of such schemes is still too high to implement practically. Moreover, communication and location sensing are usually non-cooperative, which compete mutually, e.g., in time resources \cite{wangrui2021joint}. To improve the spectrum efficiency, it is promising to realize high-accuracy location sensing using communication signals with the aid of mmWave and MIMO technologies such that communication and location sensing can be conducted on the same time-frequency resources. Few works have considered the above issues, such as \cite{liu2022sensing,yzy_TSP}. Nevertheless, it only considered to use the location sensing to improve the communication performance rather than improve the JC\&S performance, which has not exploited the full potential of the RIS-aided ISAC systems. Actually, location sensing can not only act as a service but also provide prior information for further improvement of both sensing and communication performances, which implies that it would be mutually beneficial if they can function in a cooperative mode. 

Motivated by the above issues, in this paper, we propose an RIS-aided MIMO ISAC framework, where a distributed RIS, consisting of a large-scale reflecting sub-surface and two small-scale sensing sub-surfaces, is deployed to realize simultaneous communication and location sensing for a blind-zone UE. For the proposed framework, we first design a two-phase ISAC transmission protocol, and then propose two sensing-based beamforming schemes {\color{black} to boost the performances of both communication and location sensing}, while removing the requirement of high-overhead cascaded channel estimation. Our main contributions are summarized as follows.

\begin{itemize}
    \item We propose a two-phase ISAC transmission protocol for the RIS-aided MIMO ISAC system. Specifically, each coherence block is divided into two ISAC phases. During each phase, the UE sends communication signals to the BS via the reflecting sub-surface, and simultaneously the two sensing sub-surfaces receive these signals for location sensing. \textcolor{black}{In the first phase, the communication and coarse-grained location sensing are performed concurrently by exploiting the very limited CSI knowledge, while in the second phase, by using the coarse-grained sensing information obtained from the first phase, simple-yet-efficient sensing-based beamforming designs are proposed to realize both higher-rate communication and fine-grained location sensing.}
    \item \textcolor{black}{By invoking the sensed UE's location, we propose two sensing-based beamforming schemes, i.e., the sensing-based semi-definite relaxation (S-SDR) beamforming scheme and the low-complexity sensing-based multi-beam steering (S-MBS) beamforming scheme}, to maximize the joint performance of communication and location sensing, which provide a favorable communication-sensing trade-off and meanwhile avoid the intolerable overhead of the high-dimensional cascaded channel estimation.
    \item Simulation results demonstrate the advantage of the proposed RIS-aided ISAC system over the traditional RIS-aided communication system and the effectiveness of the proposed sensing-based beamforming schemes in optimizing the JC\&S performance and making the communication-sensing trade-off. Specifically, with the proposed sensing-based beamforming schemes, the RIS-aided ISAC system can obtain almost the same communication performance as the RIS-aided communication system with perfect CSI, \textcolor{black}{while providing up to millimeter-level positioning accuracy. In addition, we show how the communication and sensing performance can be flexibly balanced and simultaneously boosted through our proposed sensing-based beamforming designs.}
\end{itemize}

The remainder of this paper is organized as follows. Section II introduces the system model of the RIS-aided MIMO ISAC system. Section III presents the corresponding location sensing algorithm. Section IV presents the sensing-based beamforming algorithms to maximize the joint performance as well as balance the sensing and communication performance. Numerical results are provided in Section V. Finally, Section VI concludes this paper.

\emph{{Notations:}} Vectors and matrices are denoted by boldface lower case and boldface upper case, respectively. The superscripts $(\cdot) ^{\text{T}}$, $( \cdot ) ^{\text{H}}$ ,$( \cdot ) ^{-1}$ and $( \cdot ) ^*$ denote  the operations of transpose, Hermitian transpose, inverse and conjugate, respectively. The Euclidean norm, absolute value, Hadamard product and Kronecker product are respectively denoted by $\left\| \cdot \right\|$, $\left| \cdot \right|$, $\odot$ and $\otimes $. $\mathbb{E} \left\{ \cdot \right\} $ denotes the statistical expectation. Moreover, $U(0,1)$ and $\mathcal{C} \mathcal{N} ( 0,\sigma ^2 ) $ denotes standard continuous uniform distribution as well as the circularly symmetric complex Gaussian (CSCG) distribution with zero mean  and variance $\sigma ^2$, respectively. For matrices,
$\left[ \cdot \right] _{ij}$ denotes the $(i,j)$-th element of a matrix, $\text{tr}( \cdot ) $ represents the matrix trace, and $\operatorname{diag}( \cdot ) $ denotes a square diagonal matrix with the elements in $( \cdot ) $ on its main diagonal. For vectors, $\left[ \cdot \right] _i$ denotes the $i$-th entry of a vector. Besides, $j$ in $e^{j\theta}$ denotes the imaginary unit. 

\vspace{-5mm} \section{System Model}
\begin{figure} [htb]
  \centering  
  \setlength{\abovecaptionskip}{0.cm}
  \includegraphics[width=0.8\linewidth]{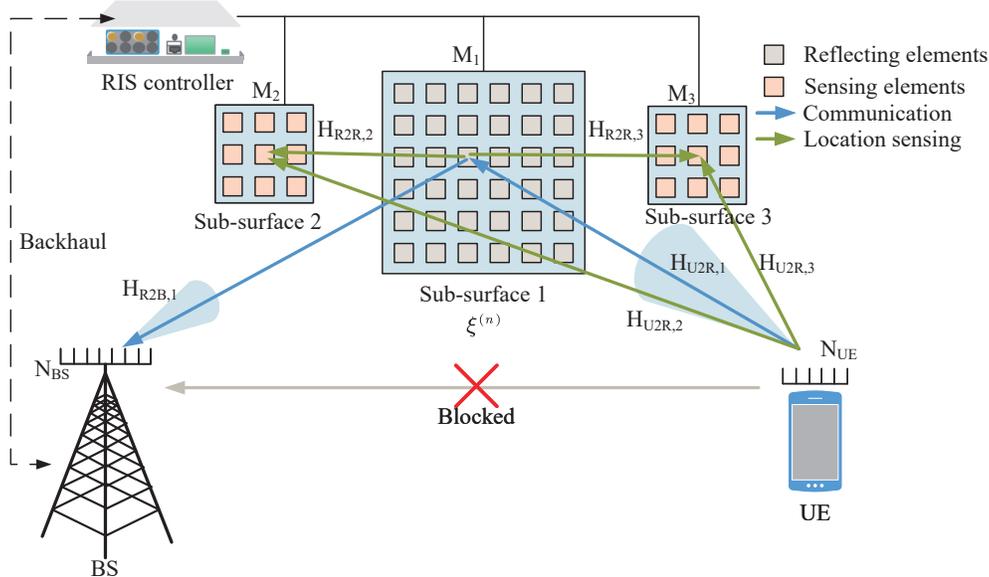}  
  \caption{Proposed RIS-aided ISAC system.}  
   \label{model scenario}
\end{figure}

As shown in Fig. \ref{model scenario}, we consider an RIS-aided system operating in the mmWave band, which consists of a BS, a UE and a distributed RIS. Due to the blockage or the unfavorable propagation environment, the direct link between the BS and the UE does not exist. Hence, a distributed RIS is deployed to assist the concurrent communication and location sensing for the blind-zone UE. The distributed RIS consists of 3 sub-surfaces. The first sub-surface (i.e., sub-surface 1) is composed of $M_{1}$ reflecting elements, while the $i$-th $(i=2,3)$ sub-surface (i.e.,  sub-surfaces 2 and 3) is composed of $M_{i}=M_{\text{s}}\ll M_{1}$ sensing elements which are able to receive the incident signals. Specifically, the reflecting sub-surface, i.e., the first sub-surface, assists the communication task by reflecting signals from the UE to the BS, \textcolor{black}{and meanwhile the two sensing sub-surfaces, i.e., the second and third sub-surfaces, which acts as the positioning reference points, carry out angle-of-arrival (AoA) estimation for obtaining the UE's location.} The BS and the UE are equipped with uniform linear arrays (ULAs) along the $y$ axis respectively, while the $i$-th sub-surface is equipped an $M_{y, i} \times M_{z, i}$ $(i=1,2,3)$ uniform rectangular array (URA) lying on the $y$-$o$-$z$ plane. Moreover, there is a backhaul link for information exchange between the BS and the RIS. In this paper, the quasi-static block-fading channel is considered for both the UE-RIS and RIS-BS channels, which remain nearly unchanged in each coherence block but vary from one block to another. Let $\mathbf{H}_{\text{U2R},i}\in \mathbb{C}^{M_{i} \times N_{\text{UE}}}$, $\mathbf{H}_{\text{R2B},1}\in \mathbb{C}^{N_{\text{BS}} \times M_{1}}$ and $\mathbf{H}_{\text{R2R},i}\in \mathbb{C}^{M_{i}\times M_{1}}$ denote the channels from the UE to the $i$-th sub-surface $(i=1,2,3)$, from the first sub-surface to the BS, and from the first sub-surface to the $i$-th sub-surface $(i=2,3)$, respectively.

\subsection{ISAC Transmission Protocol}
\begin{figure} 
  \centering  
  \includegraphics[width=11cm]{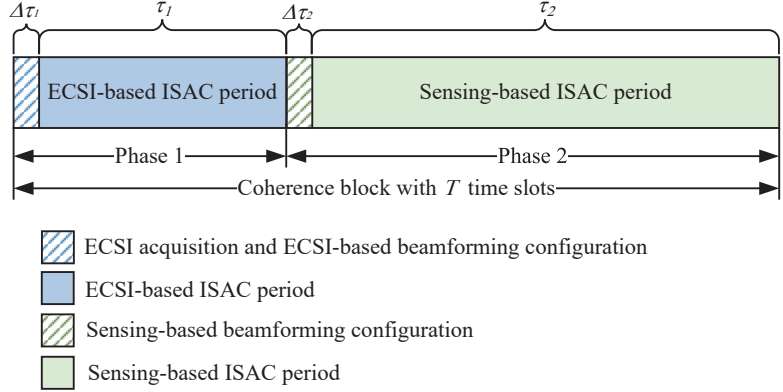}
  \caption{Proposed two-phase ISAC transmission protocol.}
  \label{transmission protocol}
\end{figure}

As shown in Fig. \ref{transmission protocol}, we focus on the ISAC transmission within one coherence block. Each coherence block consisting of $T$ time slots (symbol duration) is composed of two ISAC phases, during each of which, {\color{black} the UE sends information-carrying signals to the BS via the reflecting sub-surface and simultaneously the remaining two sensing sub-surfaces conduct location sensing by receiving these communication signals.} 

Phase 1 consists of $\Delta\tau_{1}+\tau_{1}$ time slots. \textcolor{black}{In the first $\Delta\tau_{1}$ time slots of phase 1, the BS estimates the low-dimensional effective UE-BS channel $\mathbf{H}_{\text{EU2B}}^{(1)}(\boldsymbol{\xi}^{(1)})$ for the given passive beamforming vector $\boldsymbol{\xi}^{(1)}$ via uplink pilot training, where} $\mathbf{H}_{\text{EU2B}}^{(1)}(\boldsymbol{\xi}^{(1)})=\mathbf{H}_{\text{R2B},1}\operatorname{diag}(\boldsymbol{\xi}^{(1)})\mathbf{H}_{\text{U2R},1}$ and $\boldsymbol{\xi}^{(n)}=[e^{j \vartheta_{1}^{(n)}}, \ldots, e^{j \vartheta_{m}^{(n)}}, \ldots, e^{j \vartheta_{M_{1}}^{(n)}}]^{\text{T}}$ is the passive beamforming vector of the reflecting sub-surface in the $n$-th phase ($n=1,2$). Due to the unavailability of any CSI knowledge, the passive beamforming vector $\boldsymbol{\xi}^{(1)}$ is randomly generated and remains unchanged during phase 1. 
Based on the effective channel state information (ECSI), i.e., $\mathbf{H}_{\text{EU2B}}^{(1)}(\boldsymbol{\xi}^{(1)})$, the BS optimizes active beamforming of both the BS and the UE to maximize the communication performance, and then sends the optimized beamforming results to the UE. These ECSI-based beamforming parameters will be configured at the BS and the UE in the subsequent $\tau_{1}$ time slots, where uplink communication and coarse-grained location sensing at the two sensing sub-surfaces are conducted concurrently. After obtaining the coarse-grained location sensing information in phase 1, the RIS sends it to the BS. By exploiting this sensing information, the BS optimizes active and passive beamforming to maximize the joint communication and sensing performance, and then sends the optimized beamforming results to the RIS and the UE. These sensing-based beamforming parameters will be configured at the BS, RIS, and UE, during the phase 2 to achieve higher-rate communication and fine-grained location sensing concurrently.{\color{black}\footnote{\color{black}Since the location information estimated via the location sensing algorithm in phase 1 is used for beamforming in phase 2, the joint communication-sensing performance achieved by the proposed sensing-based beamforming algorithm in phase 2 is affected by the location sensing algorithm. When the positioning error is in a relatively higher range, the proposed sensing-based beamforming scheme is very sensitive to the positioning error. However, as the positioning error decreases to a lower level, the proposed sensing-based beamforming scheme becomes less sensitive to the change of positioning error.}}

\begin{figure} 
  \centering  
  \includegraphics[width=14cm]{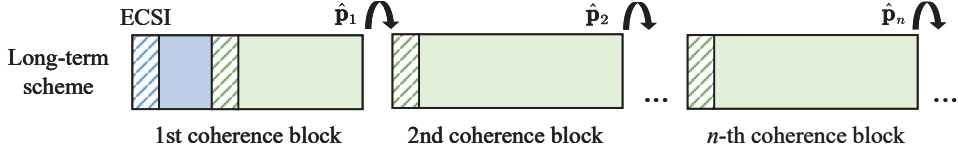}
  \caption{\color{black}The extended long-term transmission protocol.}
  \label{long_term_scheme}
\end{figure}

\begin{remark}
The main idea of the proposed transmission protocol is to exploit the coarse-grained UE's location sensed in phase 1 for improving both communication and location sensing performances in phase 2. Specifically, in phase 1, we roughly estimate the UE's location in a short time so that the joint performance in phase 2 can be quickly improved by using estimated UE's location for joint active and passive beamforming. In addition, to avoid the high pilot overhead, in phase 1, we estimate the low-dimensional effective UE-BS channel for the given passive beamforming vector instead of the high-dimensional cascaded BS-RIS-UE channel.
\end{remark}
\begin{remark}
\textcolor{black}{The proposed two-phase ISAC transmission protocol can be easily extended to the long-term transmission protocol with multiple coherence blocks. As illustrated in Fig. \ref{long_term_scheme}, for the first coherence block, we use random beamforming in the initial of the first phase due to the unavailability of any CSI knowledge in the beginning. For the subsequent coherence blocks, we remove all the phase 1's and properly design the beamforming by exploiting the fine-grained location sensing information (i.e., $\hat{\mathbf{p}}_n$) obtained in the previous coherence block.}  
\end{remark}

During phase $n$ $(n=1,2)$, the UE sends $\sqrt{\rho} s(t)$ to the BS at time slot $t \in \mathcal{N}_n$ with $\mathcal{N}_{1}=\left\{1, \cdots, \tau_{1}\right\}$ and $\mathcal{N}_{2}=\left\{\tau_{1}+1, \cdots, \tau_{1}+\tau_{2}\right\}$, where $\mathbb{E}\left\{|s(t)|^2\right\}=1$ and $\rho$ is the transmit power. The received signal at the BS via the reflecting sub-surface is
\begin{equation}
y(t)=\sqrt{\rho}\left[\mathbf{w}_{\text{BS}}^{(n)}\right]^{\text{H}} \mathbf{H}_{\text{R2B}, 1} \operatorname{diag}\left(\boldsymbol{\xi}^{(n)}\right) \mathbf{H}_{\text{U2R}, 1}\mathbf{w}_{\text{UE}}^{(n)} s(t)+\left[\mathbf{w}_{\text{BS}}^{(n)}\right]^{\text{H}} \mathbf{n}_{{\text{BS}}}(t), t \in \mathcal{N}_{n}, n=1,2,
\label{equ:ISAC Period Received Signal at BS}
\end{equation}
where $\mathbf{w}_{\text{BS}}^{(n)}$ and $\mathbf{w}_{\text{UE}}^{(n)}$, satisfying $\left\|\mathbf{w}_{\text{BS}}^{(n)}\right\|=\left\|\mathbf{w}_{\text{UE}}^{(n)}\right\|=1$, are the BS combining vector and UE beamforming vector in phase $n$, respectively. In addition, $\mathbf{n}_{\text{BS}}$ is the additive white Gaussian noise (AWGN) at the BS, whose elements follow the complex Gaussian distribution $\mathcal{C} \mathcal{N}(0, \sigma_{0}^{2})$.

The received signal at the $i$-th sub-surface $(i=2,3)$, i.e., the sensing sub-surface, is given by
\begin{equation}
\begin{aligned}
\mathbf{x}_{i}(t)=\sqrt{\rho} \mathbf{H}_{\text{U2R}, i}\mathbf{w}_{\text{UE}}^{(n)} s(t)&+\sqrt{\rho} \mathbf{H}_{\text{R2R}, i}  \operatorname{diag}\left(\boldsymbol{\xi}^{(n)}\right) \mathbf{H}_{\text{U2R}, 1} \mathbf{w}_{\text{UE}}^{(n)} s(t)\\
&+\mathbf{n}_{i}(t), i=2,3, t \in \mathcal{N}_{n}, n=1,2,
\label{equ:ISAC Period Received Signal at RIS}
\end{aligned}
\end{equation}
where $\mathbf{H}_{{\text{R2R}, i}}~(i=2,3)$ is the channel from the reflecting sub-surface to the $i$-th sub-surface. In addition, $\mathbf{n}_{i}(t)$ is the AWGN at the $i$-th sub-surface, whose elements follow the complex Gaussian distribution $\mathcal{C} \mathcal{N}(0, \sigma_{0}^{2}).$ 

The instantaneous achievable rate at time slot $t$ is given by 
\begin{equation}
R(t)=\log _{2}\left(1+\rho\left|\left[\mathbf{w}_{\text{BS}}^{(n)}\right]^{\text{H}} \mathbf{H}_{\text{R2B}, 1}  \operatorname{diag}\left(\boldsymbol{\xi}^{(n)}\right)\mathbf{H}_{\text{U2R}, 1}\mathbf{w}_{\text{UE}}^{(n)}\right|^{2}\Bigg/\sigma_{0}^{2}\right), t \in \mathcal{N}_{n}, n=1,2.
\label{equ:ISAC Period Rate}
\end{equation}

\vspace{-5mm} \subsection{Channel Model}
In general, the RIS is deployed with LoS paths to both the BS and the UE. For the considered RIS-aided ISAC system operating in the mmWave band, the NLoS path is much weaker than the LoS path. Hence, we adopt \textcolor{black}{the LoS channel model \cite{channel_model_added_1}} and characterize the channels from the reflecting sub-surface (i.e., the first sub-surface) to the BS as
\begin{equation}
\mathbf{H}_{\text{R2B}, 1}=\alpha_{\text{R2B}, 1} \mathbf{a}\left(u_{ \text{R2B}, 1}^{\text{A}}\right) \mathbf{b}_{1}^{\text{H}}\left(u_\text{R2B, 1}^{\text{D}}, v_{\text{R2B}, 1}^{\text{D}}\right),
\label{equ:Channel R2B,i}
\end{equation}
where $\alpha_{\text{R2B}, 1}$ is the complex channel gain, $\mathbf{a}$ and $\mathbf{b}_{1}$ are the array response vectors for the BS and the reflecting sub-surface, respectively. We define the two effective angles of departure (AoDs) at the reflecting sub-surface as $u_{\text{R2B}, 1}^{\text{D}}=2 \pi \frac{d_{\text{RIS}}}{\lambda} \cos (\gamma_{\text{R2B}, 1}^{\text{D}}) \sin (\varphi_{\text{R2B}, 1}^{\text{D}})$, $v_{\text{R2B}, 1}^{\text{D}}=2 \pi \frac{d_{\text{RIS}}}{\lambda} \sin (\gamma_{\text{R2B}, 1}^{\text{D}})$, where $d_{\text{RIS}}$ is the distance between two adjacent reflecting elements, $\lambda$ is the wavelength, $\gamma_{ \text{R2B}, 1}^{\text{D}}$ and $\varphi_{\text{R2B}, 1}^{\text{D}}$ are the elevation and azimuth AoDs for the link from the reflecting sub-surface to the BS, respectively. And the effective AoA at the BS is defined as $u_{\text{R2B}, 1}^{\text{A}}=2 \pi \frac{d_{\text{BS}}}{\lambda} \sin (\theta_{\text{R2B}, 1}^{\text{A}})$, where $d_{\text{BS}}$ is the distance between two adjacent antennas, and $\theta_{\text{R2B}, 1}^{\text{A}}$ is the AoA at the BS.

Similarly, \textcolor{black}{we characterize the channels from the UE to the $i$-th $(i=1,2,3)$ sub-surface as}
\begin{equation}
\mathbf{H}_{\text{U2R}, i}=\alpha_{\text{U2R}, i} \mathbf{b}_{i}\left(u_{\text{U2R}, i}^{\text{A}}, v_{\text{U2R}, i}^{\text{A}}\right)\mathbf{c}^{\text{H}}\left(u_{\text{U2R}, i}^{\text{D}}\right), i=1,2,3,
\label{equ:Channel U2R,i}
\end{equation}
where $\alpha_{\text{U2R}, i}$ is the complex channel gain, $\mathbf{b}_{i}$ are the array response vectors for the $i$-th sub-surface, $u_{\text{U2R}, i}^{\text{A}}$, $v_{\text{U2R}, i}^{\text{A}}$ and  $v_{\text{U2R}, i}^{\text{D}}$ are effective AoAs and AoDs from the UE to the $i$-th sub-surface, which are defined as $u_{\text{U2R}, i}^{\text{A}}=2 \pi \frac{d_{\text{RIS}}}{\lambda} \cos (\gamma_{\text{U2R}, i}^{\text{A}}) \sin (\varphi_{\text{U2R}, i}^{\text{A}})$, $v_{\text{U2R}, i}^{\text{A}}=2 \pi \frac{d_{\text{RIS}}}{\lambda} \sin (\gamma_{\text{U2R}, i}^{\text{A}})$, $u_{\text{U2R}, i}^{\text{D}}=2 \pi \frac{d_{\text{UE}}}{\lambda} \sin (\theta_{\text{U2R}, i}^{\text{D}})$, where $\gamma_{\text{U2R}, i}^{\text{A}}$ and $\varphi_{\text{U2R}, i}^{\text{A}}$ are the elevation and azimuth AoAs for the link from the UE to the $i$-th sub-surface, respectively, $d_{\text{UE}}$ is the distance between two adjacent antennas at the UE, and $\theta_{\text{U2R}, i}^{\text{D}}$ is the AoD at the UE. 

Also, the channel from the reflecting sub-surface to the sensing sub-surface (i.e., the $i$-th sub-surface, $i=2,3$ ) is modeled as
\begin{equation}
\mathbf{H}_{\text{R2R}, i}=\alpha_{\text{R2R}, i} \mathbf{b}_{i}\left(u_{\text{R2R}, i}^{\text{A}}, v_{\text{R2R}, i}^{\text{A}}\right) \mathbf{b}_{1}^{\text{H}}\left(u_{\text{R2R}, i}^{\text{D}}, v_{\text{R2R}, i}^{\text{D}}\right), i=2,3,
\label{equ:channel R2R,i}
\end{equation}
where $\alpha_{\text{R2R}, i}$ is the complex channel gain, and the two effective AoAs at the $i$-th sub-surface are defined as $u_{\text{R2R}, i}^{\text{A}}=2 \pi \frac{d_{\text{RIS}}}{\lambda} \cos (\gamma_{\text{R2R}, i}^{\text{A}}) \sin (\varphi_{\text{R2R}, i}^{\text{A}})$,  
$v_{\text{R2R}, i}^{\text{A}}=2 \pi \frac{d_{\text{RIS}}}{\lambda} \sin (\gamma_{\text{R2R}, i}^{\text{A}})$, where $\gamma_{\text{R2R}, i}^{\text{A}}$ and $\varphi_{\text{R2R}, i}^{\text{A}}$ are the elevation and azimuth AoAs from the reflecting sub-surface to the $i$-th sub-surface, respectively. And the two effective AoDs at the first sub-surface are defined as $u_{\text{R2R}, i}^{\text{D}}=2 \pi \frac{d_{\text{RIS}}}{\lambda} \cos (\gamma_{\text{R2R}, i}^{\text{D}}) \sin (\varphi_{\text{R2R}, i}^{\text{D}})$, 
$v_{\text{R2R}, i}^{\text{D}}=2 \pi \frac{d_{\text{RIS}}}{\lambda} \sin (\gamma_{\text{R2R}, i}^{\text{D}})$, where $\gamma_{\text{R2R}, i}^{\text{D}}$ and $\varphi_{\text{R2R}, i}^{\text{D}}$ are the elevation and azimuth AoDs from the reflecting sub-surface to the $i$-th sub-surface, respectively.

Furthermore, we assume that $d_{\text{BS}}=d_{\text{RIS}}=d_{\text{UE}}=\frac{\lambda}{2}$. Then, the array response vectors for the BS, the $i$-th sub-surface and the UE are respectively given by
\begin{align}
&\mathbf{a}(u)=\left[1, \cdots, e^{j(n-1) u}, \cdots, e^{j(N_{\text{BS}}-1) u}\right]^{\text{T}}, \\
&\mathbf{b}_{i}(u, v)=\left[1, \cdots, e^{j(n-1) u}, \cdots, e^{j\left(M_{y, i}-1\right) u}\right]^{\text{T}} \otimes\left[1, \cdots, e^{j(m-1) v}, \cdots, e^{j\left(M_{z, i}-1\right) v}\right]^{\text{T}},\\
&\label{channel_response_vector_c}\mathbf{c}(u)=\left[1, \cdots, e^{j(n-1) u}, \cdots, e^{j(N_{\text{UE}}-1) u}\right]^{\text{T}}.
\end{align}

\vspace{-5mm} \section{LOCATION SENSING} \label{subsection 3}
During both phase 1 and phase 2, the two sensing sub-surfaces carry out UE's localization via receiving the communication signals sent from the UE to the BS. The received signal at the $i$-th $(i \in\{2,3\})$ sub-surface during the $n$-th phase is expressed in (\ref{equ:ISAC Period Received Signal at RIS}), where the first term is the desired signal from the UE, which contains the UE's location information, and the second term is the interference signal from the reflecting sub-surface.

By using the total least square (TLS) \textcolor{black}{estimation of signal parameters via rotational invariance technique (ESPRIT) method \cite{ESPRIT_REFERENCE}} to process $\mathbf{x}_{i}(t)$'s, we separately estimate the effective AOAs corresponding to the $y$ axis (i.e., $u_{\text{U2R}, i}^{\text{A}}$ and $u_{\text{R2R}, i}^{\text{A}}$) and the $z$ axis (i.e., $v_{\text{U2R}, i}^{\text{A}}$ and $v_{\text{R2R}, i}^{\text{A}}$). Then, by invoking the \textcolor{black}{multiple signal classification (MUSIC) method \cite{MUSIC_REFERENCE}}, we pair the effective AOAs corresponding to the $y$ axis with those corresponding to the $z$ axis, and obtain the two AoA pairs 
$\{(u_{\text{U2R},i}^{\text{A}},v_{\text{U2R},i}^{\text{A}}), (u_{\text{R2R},i}^{\text{A}},v_{\text{R2R},i}^{\text{A}}) \}$ corresponding to each sensing sub-surface. After removing the AoA pair $(u_{\text{R2R},i}^{\text{A}},v_{\text{R2R},i}^{\text{A}})$ by exploiting sub-surfaces' locations, we obtain the AoA pair $ (u_{\text{U2R},i}^{\text{A}},v_{\text{U2R},i}^{\text{A}})$. Finally, combining the two AoA pairs $ (u_{\text{U2R},i}^{\text{A}},v_{\text{U2R},i}^{\text{A}}), i=2,3$ from the UE to the two sensing sub-surface yields the UE's location.

\begin{figure}[htbp]
\centering
\subfigure[Example of micro-surface.]{\includegraphics[width=0.3\textwidth]{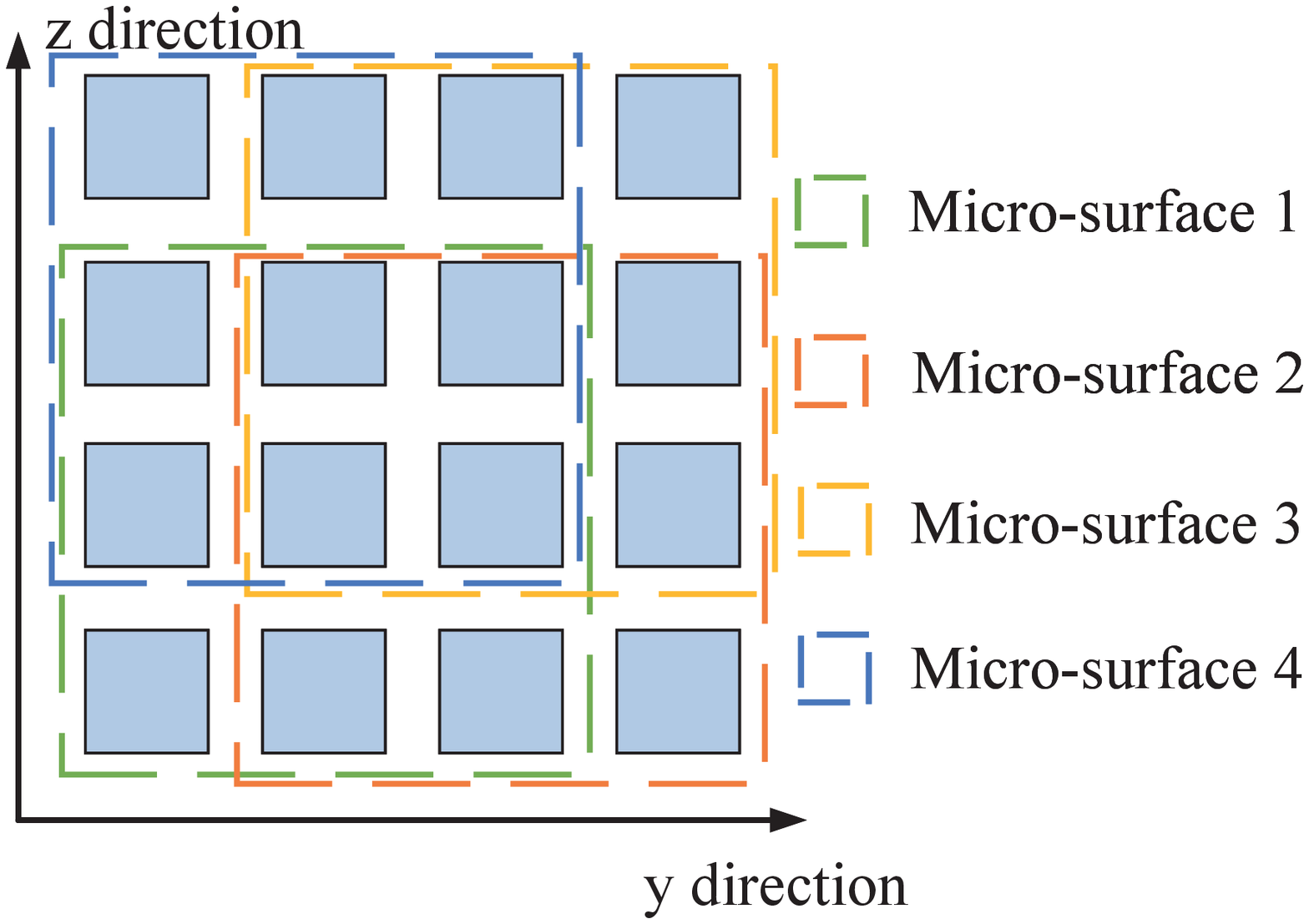} }
\subfigure[Example of auxiliary sub-surface 1.]{\includegraphics[width=0.3\textwidth]{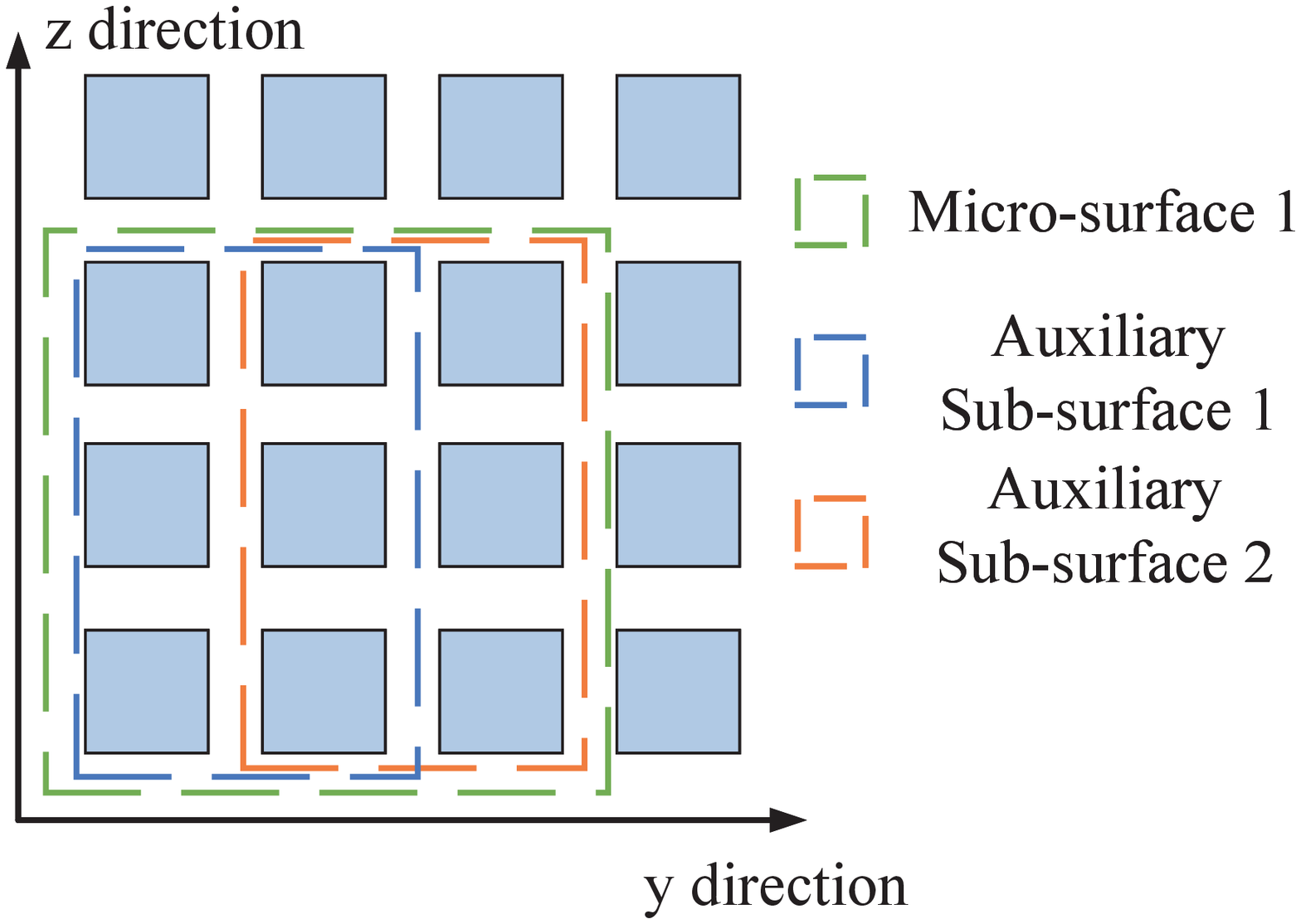} }
\subfigure[Example of auxiliary sub-surface 2.]{\includegraphics[width=0.3\textwidth]{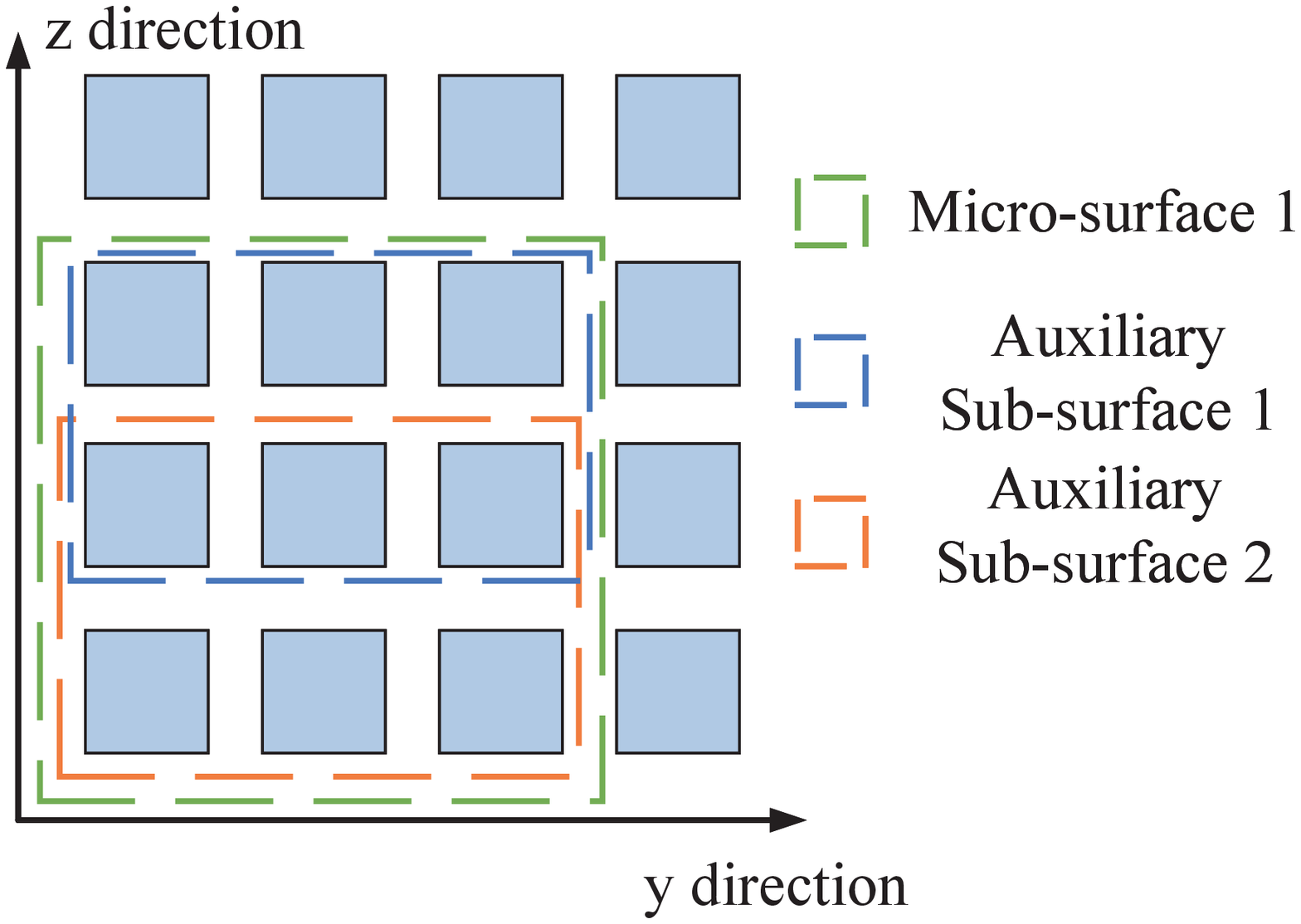} }
\caption{Illustration of the FBSS process and the auxiliary sub-surface.}
\label{FBSS and AUX illustration}
\end{figure}

Without loss of generality, we focus on the estimation of the effective AoAs at the $i$-th sub-surface in the $n$-th phase. To remove the coherency of the received signals, we adopt the forward-backward spatial smoothing (FBSS) technique \cite{Pillai1989FBSS} to preprocess the signals received at the sensing sub-surfaces. Specifically, for the $i$-th sub-surface, we construct a set of $N_{\text {micro }, i}$ micro-surfaces, each of which consists of $L_{\text {micro }, i}=Q_{y, i} \times Q_{z, i}$ sensing elements as shown in Fig. \ref{FBSS and AUX illustration}(a). Each micro-surface is shifted by one row along the $z$ direction or one column along the $y$ direction from the preceding micro-surface. We provide an example in Fig. \ref{FBSS and AUX illustration}, where a set of 4 micro-surfaces each consisting of $3 \times 3$ sensing elements are constructed.

Denote the received signals at the $m$-th micro-surface of the $i$-th sub-surface during the $n$-th phase by $\mathbf{x}_{i, m}(t) \in \mathbb{C}^{L_{\text {micro},i} \times 1}, t \in \mathcal{N}_{n}$. Following the FBSS technique, we estimate the auto-correlation matrix of $\mathbf{x}_{i, m}(t)$, i.e., $\mathbf{R}_{i}^{(n)} \triangleq \mathbb{E}\left\{\mathbf{x}_{i, m}(t)[\mathbf{x}_{i, m}(t)]^{\text{H}}\right\}$ as
\begin{align}
    &\hat{\mathbf{R}}_{i}^{(1)}=\frac{1}{2 \tau_{1} N_{\text {micro }, i}} \sum_{t \in \mathcal{N}_{1}} \sum_{m=1}^{N_{\text {micro, } i}}\left\{\mathbf{x}_{i, m}(t)\left[\mathbf{x}_{i, m}(t)\right]^{\text{H}}+\mathbf{J}\left[\mathbf{x}_{i, m}(t)\right]^{*}\left[\mathbf{x}_{i, m}(t)\right]^{\text{T}} \mathbf{J}\right\},\label{auto-correlation-tb1}\\
    &\hat{\mathbf{R}}_{i}^{(2)}=\frac{1}{2 T N_{\text {micro }, i}} \sum_{t \in {\{\mathcal{N}_{1}}\cup \mathcal{N}_{2}\}} \sum_{m=1}^{N_{\text {micro, } i}}\left\{\mathbf{x}_{i, m}(t)\left[\mathbf{x}_{i, m}(t)\right]^{\text{H}}+\mathbf{J}\left[\mathbf{x}_{i, m}(t)\right]^{*}\left[\mathbf{x}_{i, m}(t)\right]^{\text{T}} \mathbf{J}\right\},\label{auto-correlation-tb2}
\end{align}
where $\mathbf{J}$ is the exchange matrix, with the 1 elements residing on its counterdiagonal and all other elements being zero. Then, we perform eigenvalue decomposition of $\hat{\mathbf{R}}_{i}^{(n)}$ and obtain
\begin{equation}
\hat{\mathbf{R}}_{i}^{(n)}=\mathbf{U}_{i}^{(n)} \operatorname{diag}\left(\lambda_{i, 1}^{(n)}, \ldots, \lambda_{i, L_{\text {micro, }}}^{(n)}\right)\left[\mathbf{U}_{i}^{(n)}\right]^{\text{H}},\label{eigen decomposition of auto-correlation}
\end{equation}
where $\mathbf{U}_{i}^{(n)} \triangleq[\mathbf{u}_{i, 1}^{(n)}, \ldots, \mathbf{u}_{i, L_{\text {micro}}, i}^{(n)}]$ is the matrix composed of the eigenvectors and the eigenvalues $\lambda_{i, 1}^{(n)}, \ldots, \lambda_{i, L_{\text {micro}, i}}^{(n)}$ are in descending order.

\subsubsection{Estimate $u_{\rm{U2R},i}^{A}$ and $u_{\rm{R2R}, i}^{A}$ by invoking the TLS-ESPRIT algorithm}\label{sec3_subsec_1} For the $i$-th sub-surface, we construct two auxiliary sub-surfaces of its first micro-surface, each of which has the size of $L_{\text{aux}, i}=\left(Q_{y, i}-\right.$ 1) $\times Q_{z, i}$, as illustrated in Fig. \ref{FBSS and AUX illustration}(b). Then, we express the signal sub-space corresponding to the two auxiliary sub-surfaces as $\mathbf{U}_{\text{S}, i k}^{(n)} \triangleq \mathbf{J}_{k} \mathbf{U}_{\text{S}, i}^{(n)}, k=1,2$, where $\mathbf{U}_{\text{S}, i}^{(n)} \triangleq[\mathbf{u}_{i, 1}^{(n)}, \mathbf{u}_{i, 2}^{(n)}] \in \mathbb{C}^{L_{\text {micro}, i} \times 2}$ is composed of the two eigenvectors corresponding to the largest two eigenvalues of $\hat{\bf R}_{i}^{(n)}$,  and $\mathbf{J}_{k} \in \mathbb{R}^{L_{\text {aux}, i} \times L_{\text {micro}, i}}(k \in\{1,2\})$ is a selecting matrix. If the $j$-th reflecting element of the micro-surface 1 corresponds to the $i$-th element of the auxiliary sub-surface $k \in\{1,2\}$, then $\left[\mathbf{J}_{k}\right]_{i j}$ is set to be $1$. Otherwise, $\left[\mathbf{J}_{k}\right]_{i j}$ is set to be $0$.

Then, define 
\begin{equation}
\begin{aligned}\label{location_sensing_C_equation}
\mathbf{C}_{i}^{(n)} &\triangleq\left[\mathbf{U}_{\text{S}, i 1}^{(n)}, \mathbf{U}_{\text{S}, i 2}^{(n)}\right]^{\text{H}}\left[\mathbf{U}_{\text{S}, i 1}^{(n)}, \mathbf{U}_{\text{S}, i 2}^{(n)}\right] 
=\left[\begin{array}{cc}
\mathbf{V}_{i, 11}^{(n)} & \mathbf{V}_{i, 12}^{(n)} \\
\mathbf{V}_{i, 21}^{(n)} & \mathbf{V}_{i, 22}^{(n)}
\end{array}\right] \boldsymbol{\Lambda}_{C, i}^{(n)}\left[\begin{array}{cc}
\mathbf{V}_{i, 11}^{(n)} & \mathbf{V}_{i, 12}^{(n)} \\
\mathbf{V}_{i, 21}^{(n)} & \mathbf{V}_{i, 22}^{(n)}
\end{array}\right]^{\text{H}},
\end{aligned}
\end{equation}
where $\boldsymbol{\Lambda}_{C, i}^{(n)} \triangleq \operatorname{diag}(\lambda_{\text{C}, i 1}^{(n)}, \ldots, \lambda_{\text{C}, i 4}^{(n)})$ with the eigenvalues in a decreasing order, $\mathbf{V}_{i, 12}^{(n)}$ and $\mathbf{V}_{i, 22}^{(n)}$ are $2 \times 2$ matrices defined by the eigendecomposition of $\mathbf{C}_{i}^{(n)}$.

Calculate
$
\boldsymbol\Phi_{\text{TLS}, i}^{(n)}=-\mathbf{V}_{i, 12}^{(n)}[\mathbf{V}_{i, 22}^{(n)}]^{-1}
$,
perform eigendecomposition of $\boldsymbol{\Phi}_{\text{TLS}, i}^{(n)}$ and obtain its two eigenvalues $\lambda_{\text{TLS}, i l}^{(n)}, l=1,2$, based on which, we estimate the two effective AoAs at the $i$-th sub-surface regarding to the $y$ axis as
$
\check{u}_{il}^{(n)}=\operatorname{angle}\left(\lambda_{\text{TLS}, i l}^{(n)}\right),~l=1,2.
$

It is worth noting that the estimators of $u_{\text{U2R}, i}^{\text{A}}$ and $u_{\text{R2R}, i}^{\text{A}}$ in the $n$-th phase belong to the set $\mathcal{U}_{i}^{(n)} \triangleq \{\check{u}_{il}^{(n)}, l=1,2\}$, i.e., $\hat{u}_{\text{U2R}, i}^{\text{A},(n)}, \hat{u}_{\text{R2R}, i}^{\text{A},(n)} \in \mathcal{U}_{i}^{(n)}$.

\subsubsection{Estimate $v_{\rm{U2R}, i}^{A}$ and $v_{\rm{R2R}, i}^{A}$ by exploiting the TLS-ESPRIT algorithm}\label{sec3_subsec_2} For the $i$-th sub-surface, we construct two auxiliary sub-surfaces of its first micro-surface, each consisting of $\tilde{L}_{\text {aux }, i}=Q_{y, i} \times\left(Q_{z, i}-1\right)$ sensing elements, as illustrated in Fig. \ref{FBSS and AUX illustration}(b). Following the similar process of estimating $u_{\text{U2R}, i}^{\text{A}}$ and $u_{\text{R2R}, i}^{\text{A}}$, we obtain the two effective AoAs at the $i$-th sub-surface corresponding to the $z$ axis as $\check{v}_{i l}^{(n)}, l=1,2$. And the estimators of $v_{\text{U2R}, i}^{\text{A}}$ and $v_{\text{R2R}, i}^{\text{A}}$ in the $n$-th phase belong to the set $\mathcal{V}_{i}^{(n)} \triangleq\{\check{v}_{il}^{(n)}, l=1,2\}$, i.e., $\hat{v}_{\text{U2R, i}}^{\text{A},(n)}, \hat{v}_{\text{R2R},i}^{\text{A},(n)} \in \mathcal{V}_{i}^{(n)}$.

\subsubsection{Pair $\check{u}_{i l}^{(n)}$ and $\check{v}_{i l}^{(n)}$ by using the MUSIC algorithm}\label{sec3_subsec_3} Let $\check{f}_{i, l s}^{(n)} \triangleq \mathbf{b}_{\text {micro}, i}^{\text{H}}(\check{u}_{i l}^{(n)}, \check{v}_{i s}^{(n)}) \mathbf{U}_{\text{N}, i}^{(n)}$, where $\mathbf{U}_{\text{N}, i}^{(n)} \triangleq[\mathbf{u}_{i, 3}^{(n)}, \ldots, \mathbf{u}_{i, L_{\text {micro }, i}}^{(n)}] \in \mathbb{C}^{L_{\text {micro }, i} \times(L_{\text {micro}, i}-2)}$ consists of  the  eigenvectors corresponding to the smallest $L_{\text{micro},i}-2$  eigenvalues of $\hat{\bf R}_{i}^{(n)}$, and $\mathbf{b}_{\text {micro}, i}$ is the array response vector of the micro-surface on the $i$-th sub-surface. Then we compute $f(\check{u}_{i l}^{(n)}, \check{v}_{i s}^{(n)})=\check{f}_{i, l s}^{(n)}[\check{f}_{i, l s}^{(n)}]^{\text{H}}, l, s=1,2,$ and choose the two minima $f(\hat{u}_{i l}^{(n)}, \hat{v}_{i l}^{(n)}), l=1,2$, where $\hat{u}_{i l}^{(n)} \in \mathcal{U}_{i}^{(n)}$ and $\hat{v}_{i l}^{(n)} \in \mathcal{V}_{i}^{(n)} .$ As such, we obtain two AoA pairs $(\hat{u}_{i l}^{(n)}, \hat{v}_{i l}^{(n)}), l=1,2$.

\subsubsection{Determine $(u_{\rm{U2R}, i}^{\rm{A}}, v_{\rm{U2R}, i}^{\text{A}})$}\label{subsubsection_4} Since there are two pairs of effective AoAs, we need to determine which pair corresponds to $(u_{\text{U2R}, i}^{\text{A}}, v_{\text{U2R}, i}^{\text{A}})$. Denote the locations of the BS and the $i$-th sub-surface by $\mathbf{q}_{\text{BS}}=(x_{\text{BS}}, y_{\text{BS}}, z_{\text{BS}})$ and $\mathbf{q}_{i}=(x_{i}, y_{i}, z_{i})$, respectively. Since the locations of the BS and all the sub-surfaces are fixed, we assume that these locations are perfectly known. By invoking the location information, we compute the effective AoAs from the first sub-surface to the $i$-th sub-surface as $u_{\text{R2R}, i}^{\text{A}}=\frac{y_{i}-y_{1}}{\left\|\mathbf{q}_{i}-\mathbf{q}_{1}\right\|}$, $v_{\text{R2R}, i}^{\text{A}}=\frac{z_{i}-z_{1}}{\left\|\mathbf{q}_{i}-\mathbf{q}_{1}\right\|}$.

After removing the effective AoA pair corresponding to $(u_{\text{R2R}, i}^{\text{A}}, v_{\text{R2R}, i}^{\text{A}})$ from $\{(\hat{u}_{i l}^{(n)}, \hat{v}_{i l}^{(n)}) \mid l=1,2\}$, we obtain the estimator for $(u_{\text{U2R}, i}^{\text{A},(n)}, v_{\text{U2R}, i}^{\text{A},(n)})$, i.e., $(\hat{u}_{\text{U2R}, i}^{\text{A},(n)}, \hat{v}_{\text{U2R}, i}^{\text{A},(n)})$.

\subsubsection{Estimate the UE's location according to the estimated effective AoA pairs}\label{sec3_subsec_5}
\textcolor{black}{Combining the locations of the two sensing sub-surfaces and the two estimated effective AoA pairs yields the UE's location  $\hat{\mathbf{q}}_{\text{UE}}^{(n)}=[\hat{x}_{\text{UE}}^{(n)}, \hat{y}_{\text{UE}}^{(n)}, \hat{z}_{\text{UE}}^{(n)}]^{\text{T}}$. } 

The detailed process of location sensing is summarized in Algorithm \ref{alg:LOCATION_SENSING}.{\color{black}\footnote{\color{black}Note that ESPRIT first decomposes the signal space into signal subspace and noise subspace, and then processes the signals in the signal subspace \cite{van2004optimum}. Therefore, after the transformation of signal space, the effective SNR corresponding to the estimation of each AoA in the signal subspace increases linearly with the number of sensing elements.}}

\begin{algorithm}[htp]
\caption{{\color{black} Location Sensing Algorithm}}
\label{alg:LOCATION_SENSING}
\begin{algorithmic}[1]
\REQUIRE $\mathbf{x}_{i}(t)$.
\STATE Adopt the FBSS technique to process the signals and obtain the received signals at the $m$-th micro-surface of the $i$-th sub-surface during the $n$-th phase, i.e. $\mathbf{x}_{i, m}(t) \in \mathbb{C}^{L_{\text {micro},i} \times 1}, t \in \mathcal{N}_{n}$. 
\STATE Compute the auto-correlation matrix of $\mathbf{x}_{i, m}(t)$ and obtain $\mathbf{R}_{i}^{(n)}$ via (\ref{auto-correlation-tb1}) and (\ref{auto-correlation-tb2}).
\STATE Perform eigenvalue of $\mathbf{R}_{i}^{(n)}$ via (\ref{eigen decomposition of auto-correlation}) and obtain $\mathbf{U}_{i}^{(n)}$.
\STATE Estimate $u_{\rm{U2R},i}^{A}$, $u_{\rm{R2R}, i}^{A}$, $v_{\rm{U2R}, i}^{A}$ and $v_{\rm{R2R}, i}^{A}$ by invoking the TLS-ESPRIT algorithm according to \ref{sec3_subsec_1} and \ref{sec3_subsec_2}, respectively.
\STATE Pair $\check{u}_{i l}^{(n)}$ and $\check{v}_{i l}^{(n)}$ by using the MUSIC algorithm according to \ref{sec3_subsec_3}.
\STATE Determine $(u_{\rm{U2R}, i}^{\rm{A}}, v_{\rm{U2R}, i}^{\text{A}})$ by using locations of the BS and sub-surfaces according to \ref{subsubsection_4}.
\STATE Estimate the UE's location $\hat{\mathbf{q}}_{\text{UE}}^{(n)}$ via the estimated effective AoA pairs according to \ref{sec3_subsec_5}.
\ENSURE The estimated UE's location $\hat{\mathbf{q}}_{\text{UE}}^{(n)}=[\hat{x}_{\text{UE}}^{(n)}, \hat{y}_{\text{UE}}^{(n)}, \hat{z}_{\text{UE}}^{(n)}]^{\text{T}}$.
    \end{algorithmic}
\end{algorithm}

\section{\textcolor{black}{SENSING-BASED BEAMFORMING SCHEMES}}\label{section_beamforming}
In phase 1, by exploiting the low-dimensional ECSI of the UE-BS effective channel, we adopt the maximal ratio transmission (MRT) at the UE and maximal ratio combining (MRC) at the BS to maximize the received signal-to-noise ratio (SNR) of the BS, while realizing coarse-grained location sensing. Then we use the \textcolor{black}{coarse-grained UE's location sensed in phase 1 for beamforming in phase 2, with both communication and sensing performances considered.}

\vspace{-3mm} \subsection{ECSI-based Beamforming in Phase 1}
The passive beamforming vector ${\boldsymbol{\xi}}^{(1)}$ is generated randomly and remains unchanged during the first phase, and thus only the active beamforming of the BS and the UE needs to be optimized. By exploiting the ECSI $\mathbf{H}_{\text{EU2B}}^{(1)}(\boldsymbol{\xi}^{(1)})=\mathbf{H}_{\text{R2B},1}\operatorname{diag}(\boldsymbol{\xi}^{(1)})\mathbf{H}_{\text{U2R},1}$
for the given passive beamforming vector $\boldsymbol{\xi}^{(1)}$, we adopt the MRT-MRC method to maximize the received SNR of the BS via designing the beamforming vectors of the BS and the UE. Note that the received signal at the BS via the reflecting sub-surface in phase 1 can be expressed as 
$
    y(t)=\sqrt{\rho}\left[\mathbf{w}_{\text{BS}}^{(1)}\right]^{\text{H}} \mathbf{H}_{\text{EU2B}}^{(1)}\mathbf{w}_{\text{UE}}^{(1)} s(t)+\left[\mathbf{w}_{\text{BS}}^{(1)}\right]^{\text{H}} \mathbf{n}_{\text{BS}}(t), t \in \mathcal{N}_{1}.
$ \textcolor{black}{By invoking the results in \cite{lo1999maximum}, we obtain the optimal solution to the above problem as $\mathbf{w}_{\text{UE}}^{(1)\star}=\mathbf{w}_{\text{MRT}},~\mathbf{w}_{\text{BS}}^{(1)\star}=\frac{\mathbf{H}_{\text{EU2B}^{(1)}}\mathbf{w}_{\text{MRT}}}{\left\|\mathbf{H}_{\text{EU2B}^{(1)}}\mathbf{w}_{\text{MRT}}\right\|}$, where $\mathbf{w}_{\text{MRT}}$ is the unit eigenvector corresponding to the maximum eigenvalue of the matrix ${[\mathbf{H}_{\text{EU2B}}^{(1)}]}^{\text{H}}\mathbf{H}_{\text{EU2B}}^{(1)}$.}

\vspace{-3mm} \subsection{Sensing-based Beamforming in Phase 2}
In phase 2, we use the UE's location sensed in phase 1 for joint active and passive beamforming to improve both communication and location sensing performances concurrently. 

The received signal at the sensing sub-surface in phase 2 can be expressed as
\begin{equation}\label{equ:sensing sub-surface received signal in phase 2}
\begin{aligned}
\mathbf{x}_{i}(t) &=\underbrace{\sqrt{\rho} \mathbf{H}_{\text{U2R}, i}\mathbf{w}_{\text{UE}}^{(2)} s(t)}_{\text{UE-RIS direct link}}+\underbrace{\sqrt{\rho} \mathbf{H}_{\text{R2R}, i}  \operatorname{diag}\left(\boldsymbol{\xi}^{(2)}\right) \mathbf{H}_{\text{U2R}, 1} \mathbf{w}_{\text{UE}}^{(2)} s(t)}_{\text{UE-RIS reflecting link}}+\mathbf{n}_{i}(t),i=2,3, t \in \mathcal{N}_{2},
\end{aligned}
\end{equation}
where the first term is the desired signal directly from the UE to the sensing sub-surface, while the second term is the interference signal reflected by the reflecting sub-surface.

According to the AoA estimation theory, the accuracy of AoA estimation using TLS-ESPRIT is related to the number of snapshots (i.e., the time of sensing), the size of the array, and the received SNR \cite{van2004optimum}. Hence, for the fixed sensing time and array size, the accuracy improves with the received SNR. Moreover, since the location sensing scheme proposed in Subsection \ref{subsection 3} is able to effectively eliminate the unwanted interference signal from the reflecting sub-surface, the impact of interference signal reflected by the reflecting sub-surface is trivial to the sensing accuracy. Hence, the accuracy is determined by the received SNR corresponding to the first term. As such, we adopt the sensing SNR, i.e., $SNR_{\text{sen}}$, as the performance evaluation metric for sensing and define it as
\begin{equation}\label{def:snr_sen}
    SNR_{\text{sen}}=\sum_{i=2,3}SNR_{i,\text{d}}=\frac{\rho}{{\sigma_0}^2}\sum_{i=2,3}\left\|\mathbf{H}_{\text{U2R},i}\mathbf{w}_{\text{UE}}^{(2)}\right\|^{2},
\end{equation}
where $SNR_{i,\text{d}}~(i=2,3)$ is the received SNR of the $i$-th sub-surface, via the UE-RIS direct link.

Similarly, we write the received signal at the BS as
\begin{equation}\label{equ:BS received signal in phase 2}
y(t)=\sqrt{\rho}\left[\mathbf{w}_{\text{BS}}^{(2)}\right]^{\text{H}} \mathbf{H}_{\text{R2B}, 1} \operatorname{diag}\left(\boldsymbol{\xi}^{(2)}\right) \mathbf{H}_{\text{U2R}, 1}\mathbf{w}_{\text{UE}}^{(2)} s(t)+\left[\mathbf{w}_{\text{BS}}^{(2)}\right]^{\text{H}} \mathbf{n}_{{\text{BS}}}(t), t \in \mathcal{N}_{2}
\end{equation}
and adopt the communication SNR, i.e., $SNR_{\text{com}}$, as the performance evaluation metric for communication, which is defined as
\begin{equation}\label{def:snr_com}
    SNR_{\text{com}}=SNR_{\text{BS}}={\frac{\rho}{{\sigma_0}^2}}\left|[\mathbf{w}_{\text{BS}}^{(2)}]^{\text{H}}\mathbf{H}_{\text{R2B},1}\operatorname{diag}\left(\boldsymbol{\xi}^{(2)}\right)\mathbf{H}_{\text{U2R},1}\mathbf{w}_{\text{UE}}^{(2)}\right|^{2},
\end{equation}
where $SNR_{\text{BS}}$ is the received SNR of the BS via the reflecting sub-surface.

Then, we propose to use the weighted sum of the sensing SNR, i.e., $SNR_{\text{sen}}$, and the communication SNR, i.e., $SNR_{\text{com}}$, as the metric for characterizing the JC\&S performance, and focus on maximizing the weighted SNR by jointly optimizing the receive beamforming vector of the BS $\mathbf{w}_{\text{BS}}^{(2)}$, the passive beamforming vector of the reflecting sub-surface $\boldsymbol{\xi}^{(2)}$, as well as the transmit beamforming vector of the UE $\mathbf{w}_{\text{UE}}^{(2)}$. Mathematically, we formulate the weighted SNR maximization problem as 
\begin{subequations}\label{P1:opt-problem in ISAC Period}
\begin{align} 
\text{(P1)}:~\max _{\mathbf{w}_{\text{BS}}^{(2)},\boldsymbol{\xi}^{(2)},\mathbf{w}_{\text{UE}}^{(2)}} 
& \frac{\varrho}{\zeta_{\text{s}}} SNR_{\text{sen}}+\frac{1-\varrho}{\zeta_{\text{c}}}SNR_{\text{\text{com}}},\\
\text { s.t. }
&\left| SNR_{2,\text{d}}-SNR_{3,\text{d}}\right|\leqslant\epsilon,\label{SNR_difference_constraint}\\
&\left\|\mathbf{w}_{\text{BS}}^{(2)}\right\|=1,\label{constraint:BS}\\
& \left|\left[\boldsymbol{\xi}^{(2)}\right]_{i}\right|=1, i=1, \cdots, M_{1},\label{constraint:RIS}\\
&\left\|\mathbf{w}_{\text{UE}}^{(2)}\right\|=1,\label{constraint:UE}
\end{align}
\end{subequations}
where $\varrho\in [0,1]$ is the trade-off factor for communication and sensing performance.\textcolor{black}{\footnote{The competitive relationship where communication and sensing compete with each other for spatial resources is reflected by adjusting the trade-off factor $\varrho$.}} $\zeta_s$ and $\zeta_c$ are respectively the scaling factors for the sensing SNR and the communication SNR to make sure that the scaled sensing SNR, i.e., $\frac{SNR_{\text{sen}}}{\zeta_{\text{s}}}$, and communication SNR, i.e., $\frac{SNR_{\text{com}}}{\zeta_{\text{c}}}$, are in approximate order of magnitude.\textcolor{black}{\footnote{By doing this, both the communication and sensing performances become more sensitive to changes in the trade-off factor $\varrho$. Otherwise, $\varrho$ may fail to adjust sensing and communication SNRs due to the great difference between $SNR_{\text{sen}}$ and $SNR_{\text{com}}$.}} \textcolor{black}{And $\epsilon~(\epsilon>0)$ represents the upper threshold of the difference between $SNR_{2,\mathrm{d}}$ and $SNR_{3,\mathrm{d}}$. The location sensing accuracy is affected by both $SNR_{2,\text{d}}$ and $SNR_{3,\text{d}}$ and mainly determined by the worse one. It can be easily proved that, maximizing $\frac{\varrho}{\zeta _{\mathrm{s}}}SNR_{\mathrm{sen}}$ in the objective function, together with the constraint (\ref{SNR_difference_constraint}), is approximately equivalent to maximizing $\min \left\{ SNR_{2,\mathrm{d}},SNR_{3,\mathrm{d}} \right\} $, when $\epsilon$ is very close to 0.} 

By substituting (\ref{def:snr_sen}) and (\ref{def:snr_com}) into (P1), the problem (P1) is equivalent to
\begin{subequations}
\begin{align} \label{problem:3}
\text{(P2)}:\max _{\mathbf{w}_{\text{BS}}^{(2)},\boldsymbol{\xi}^{(2)},\mathbf{w}_{\text{UE}}^{(2)}}
&\frac{\varrho}{\zeta_{\text{s}}} \underbrace{\sum_{i=2,3}\frac{\rho}{{\sigma_0}^2}\left\|\mathbf{H}_{\text{U2R},i}\mathbf{w}_{\text{UE}}^{(2)}\right\|^{2}}_{\text{sensing part}}\!\!+\frac{1-\varrho}{\zeta_{\text{c}}}\!\times\!\underbrace{{\frac{\rho}{{\sigma_0}^2}}\left|[\mathbf{w}_{\text{BS}}^{(2)}]^{\text{H}}\mathbf{H}_{\text{R2B},1}\operatorname{diag}\left(\boldsymbol{\xi}^{(2)}\right)\mathbf{H}_{\text{U2R},1}\mathbf{w}_{\text{UE}}^{(2)}\right|^{2}}_{\text{communication part}}\!\!,\\
\text { s.t. }
&\left| \left\|\mathbf{H}_{\text{U2R},2}\mathbf{w}_{\text{UE}}^{(2)}\right\|^{2}-\left\|\mathbf{H}_{\text{U2R},3}\mathbf{w}_{\text{UE}}^{(2)}\right\|^{2}\right|\leqslant\epsilon_{0},\label{constraint:DIFF_norm}\\
&(\ref{constraint:BS}),~(\ref{constraint:RIS}),~(\ref{constraint:UE}),\label{constraint:set}
\end{align}
\end{subequations}
where $\epsilon_{0}\triangleq \frac{\epsilon\sigma_{0}^{2}}{\rho}$.

To decouple the optimization variables $\mathbf{w}_{\text{BS}}^{(2)}$, $\boldsymbol{\xi}^{(2)}$, and $\mathbf{w}_{\text{UE}}^{(2)}$, we recast the problem (P2) as
\begin{equation}
 \label{problem:4}
\text{(P3)}:~\max _{\mathbf{w}_{\text{UE}}^{(2)}}
\left\{\frac{\varrho}{\zeta_{\text{s}}} {\widehat{SNR}_{\text{sen}}}+\frac{1-\varrho}{\zeta_{\text{c}}}\left[\max_{\mathbf{w}_{\text{BS}}^{(2)},\boldsymbol{\xi}^{(2)}}\widehat{SNR}_{\text{com}}\right]\right\},~\text { s.t. } (\ref{constraint:DIFF_norm}),~({\ref{constraint:set}}),
\end{equation}
where ${\widehat{SNR}_{\text{sen}}}\triangleq\sum_{i=2,3}\left\|\mathbf{H}_{\text{U2R},i}\mathbf{w}_{\text{UE}}^{(2)}\right\|^{2}$ and ${\widehat{SNR}_{\text{com}}}\triangleq\left|[\mathbf{w}_{\text{BS}}^{(2)}]^{\text{H}}\mathbf{H}_{\text{R2B},1}\operatorname{diag}\left(\boldsymbol{\xi}^{(2)}\right)\mathbf{H}_{\text{U2R},1}\mathbf{w}_{\text{UE}}^{(2)}\right|^{2}$.

Then, we decompose the problem (P3) equivalently into two sub-problems, i.e., the sub-problem corresponding to the beamforming vectors of the BS and the reflecting sub-surface
\begin{equation}
\label{problem:4-a}
\text{(P4-a)}:~\max_{\mathbf{w}_{\text{BS}}^{(2)},\boldsymbol{\xi}^{(2)}}\widehat{SNR}_{\text{com}},~\text { s.t. } ({\ref{constraint:BS}}),~({\ref{constraint:RIS}}),
\end{equation}
and the sub-problem corresponding to the beamforming vector of the UE
\begin{equation}
 \label{problem:4-b}
\text{(P4-b)}:~\max _{\mathbf{w}_{\text{UE}}^{(2)}}
\left\{ \frac{\varrho}{\zeta_{\text{s}}} {\widehat{SNR}_{\text{sen}}}+\frac{1-\varrho}{\zeta_{\text{c}}}{\widehat{SNR}_{\text{com}}}^{\text{max}}\right\},~\text { s.t. }({\ref{constraint:UE}}),~(\ref{constraint:DIFF_norm}),
\end{equation}
where ${\widehat{SNR}_{\text{com}}}^{\text{max}}$ is the optimal value of the objective function in the problem (P4-a).

First, by substituting (\ref{equ:Channel R2B,i}) and (\ref{equ:Channel U2R,i}) into (P4-a), we rewrite the problem (P4-a) as
\begin{subequations}
\begin{align} 
\text{(P4-a$'$)}:\!\max_{\mathbf{w}_{\text{BS}}^{(2)},\boldsymbol{\xi}^{(2)}}&\left|\alpha_{\text{com}}[\mathbf{w}_{\text{BS}}^{(2)}]^{\text{H}}\mathbf{a}\left(u_{ \text{R2B}, 1}^{\text{A}}\right) \mathbf{b}_{1}^{\text{H}}\left(u_\text{R2B,1}^{\text{D}}, v_{\text{R2B},1}^{\text{D}}\right)\operatorname{diag}\left(\boldsymbol{\xi}^{(2)}\right)\mathbf{b}_{1}\left(u_{\text{U2R}, 1}^{\text{A}}, v_{\text{U2R}, 1}^{\text{A}}\right) {q}(\mathbf{w}_{\text{UE}}^{(2)})\right|^{2}\!\!,\label{P4-a'}\\
\text { s.t. }
&({\ref{constraint:BS}}),~({\ref{constraint:RIS}}),
\end{align}
\end{subequations}
where $\alpha_{\text{com}}\triangleq\alpha_{\text{R2B}, 1}\alpha_{\text{U2R}, 1}$ and ${q}(\mathbf{w}_{\text{UE}}^{(2)})\triangleq\mathbf{c}^{\text{H}}\left(u_{\text{U2R}, 1}^{\text{D}}\right)\mathbf{w}_{\text{UE}}^{(2)}$. 
Furthermore, we decompose the problem (P4-a$'$) equivalently into two sub-problems, i.e.,
\begin{equation}
    \max _{\mathbf{w}_{\text{BS}}^{(2)}} \label{sub-problem 1 real} \left|[\mathbf{w}_{\text{BS}}^{(2)}]^{\text{H}}\mathbf{a}\left(u_{ \text{R2B}, 1}^{\text{A}}\right)\right|^2,~\text { s.t. }(\ref{constraint:UE}),
\end{equation}
and 
\begin{equation}
    \max _{\boldsymbol{\xi}^{(2)}}   \label{sub-problem 2 real}
    \left|\mathbf{b}_{1}^{\text{H}}\left(u_\text{R2B,1}^{\text{D}}, v_{\text{R2B},1}^{\text{D}}\right)\operatorname{diag}\left(\boldsymbol{\xi}^{(2)}\right)\mathbf{b}_{1}\left(u_{\text{U2R}, 1}^{\text{A}}, v_{\text{U2R}, 1}^{\text{A}}\right)\right|^2,~\text{s.t.}~(\ref{constraint:RIS}).
\end{equation}

The optimal solutions to sub-problems (\ref{sub-problem 1 real}) and (\ref{sub-problem 2 real}) are respectively given by
\begin{align}
\mathbf{w}_{\text{BS}}^{(2)\star}&=\frac{1}{\sqrt{N_{\text{BS}}}}\mathbf{a}\left(u_{ \text{R2B}, 1}^{\text{A}}\right),\label{com_result_1}\\
\boldsymbol{\xi}^{(2)\star}&=\operatorname{diag}\left(\mathbf{b}_{1}^{\ast}\left(u_{\text{U2R}, 1}^{\text{A}}, v_{\text{U2R}, 1}^{\text{A}}\right)\right) \mathbf{b}_{1}\left(u_{\text{R2B}, 1}^{\text{D}}, v_{\text{R2B}, 1}^{\text{D}}\right) \notag\\
&=\mathbf{b}_{1}^{\ast}\left(u_{\text{U2R}, 1}^{\text{A}}, v_{\text{U2R}, 1}^{\text{A}}\right) \odot\mathbf{b}_{1}\left(u_{\text{R2B}, 1}^{\text{D}}, v_{\text{R2B}, 1}^{\text{D}}\right).\label{com_result_2}
\end{align}

By substituting (\ref{com_result_1}) and (\ref{com_result_2}) into (\ref{P4-a'}), we obtain the optimal value of the objective function in the problem (P4-a$'$) as
\begin{equation}\label{hat_SNR_max_com}
{\widehat{SNR}_{\text{com}}}^{\text{max}}=\left|\alpha_{\text{com}}{\gamma}{q}(\mathbf{w}_{\text{UE}}^{(2)})\right|^{2},
\end{equation}
where $\gamma\triangleq\left[\mathbf{w}_{\text{BS}}^{(2)\star} \right]^{\text{H}} \mathbf{a}\left(u_{ \text{R2B}, 1}^{\text{A}}\right) \mathbf{b}_{1}^{\text{H}}\left(u_\text{R2B,1}^{\text{D}}, v_{\text{R2B},1}^{\text{D}}\right)\operatorname{diag}\left(\boldsymbol{\xi}^{(2)\star}\right)\mathbf{b}_{1}\left(u_{\text{U2R}, 1}^{\text{A}}, v_{\text{U2R}, 1}^{\text{A}}\right)$.

Second, we solve the problem (P4-b) using the above results. Specifically, we substitute (\ref{hat_SNR_max_com}) into the problem (P4-b) and obtain
\begin{equation}
\text{(P5)}:~\max _{\mathbf{w}_{\text{UE}}^{(2)}}
~\frac{\varrho}{\zeta_{\text{s}}} \sum_{i=2,3}\left\|\mathbf{H}_{\text{U2R},i}\mathbf{w}_{\text{UE}}^{(2)}\right\|^{2}+\frac{1-\varrho}{\zeta_{\text{c}}}\left|\alpha_{\text{com}}{\mathbf{\gamma}}{q}(\mathbf{w}_{\text{UE}}^{(2)})\right|^{2},~\text { s.t. }~(\ref{constraint:UE}),~(\ref{constraint:DIFF_norm}).
\end{equation}
Next, by substituting (\ref{equ:Channel R2B,i}) and (\ref{equ:Channel U2R,i}) into (P5), we simplify the above problem as
\begin{subequations}
\begin{align} 
\text{(P6)}:~\max _{\mathbf{w}_{\text{UE}}^{(2)}}
&~\frac{\varrho}{\zeta_{\text{s}}} \sum_{i=2,3}|\alpha_{\text{U2R}, i}|^{2}M_{s}\left| \mathbf{c}^{\text{H}}\left(u_{\text{U2R}, i}^{\text{D}}\right)\mathbf{w}_{\text{UE}}^{(2)}\right|_2^{2}+\frac{1-\varrho}{\zeta_{\text{c}}}|\alpha_{\text{com}}|^{2}{\mathbf{\gamma}}^{2}\left|{q}(\mathbf{w}_{\text{UE}}^{(2)})\right|^{2},\\
\text { s.t. }
&0\leqslant\left|\mathbf{c}^{\text{H}}\left(u_{\text{U2R},2}^{\text{D}}\right)\mathbf{w}_{\text{UE}}^{(2)}\right|^2-\kappa\left|\mathbf{c}^{\text{H}}\left(u_{\text{U2R},3}^{\text{D}}\right)\mathbf{w}_{\text{UE}}^{(2)}\right|^2\leqslant\epsilon_1, (\ref{constraint:UE}),
\end{align}
\end{subequations}
where 
\begin{equation}
    \kappa\triangleq\frac{\left|\alpha_{\text{U2R}, 3}\right|^2}{\left|\alpha_{\text{U2R}, 2}\right|^2}, \epsilon_1\triangleq\frac{\epsilon_0}{\left|\alpha_{\text{U2R}, 2}\right|^2 M_{s}}.\label{kappa_cal_and_epsilon_cal}
\end{equation}

Let ${\zeta_{\text{s}}}=\frac{1}{2}(\left|\alpha_{\text{U2R}, 2}\right|^2+\left|\alpha_{\text{U2R}, 3}\right|^2)M_{s}$ and ${\zeta_{\text{c}}}=\left|\alpha_{\text{com}}\right|^2\gamma^{2}$. We simplify the problem (P6) as
\begin{subequations}
\begin{align} 
\text{(P7)}:~\max _{\mathbf{w}_{\text{UE}}^{(2)}}
&~\varrho \sum_{i=2,3}\eta_{i}\left| \mathbf{c}^{\text{H}}\left(u_{\text{U2R}, i}^{\text{D}}\right)\mathbf{w}_{\text{UE}}^{(2)}\right|^{2}+(1-\varrho)\left|\mathbf{c}^{\text{H}}\left(u_{\text{U2R}, 1}^{\text{D}}\right)\mathbf{w}_{\text{UE}}^{(2)}\right|^{2},\label{problem:7 opt obj}\\
\text { s.t. }
&0\leqslant\left|\mathbf{c}^{\text{H}}\left(u_{\text{U2R},2}^{\text{D}}\right)\mathbf{w}_{\text{UE}}^{(2)}\right|^2-\kappa\left|\mathbf{c}^{\text{H}}\left(u_{\text{U2R},3}^{\text{D}}\right)\mathbf{w}_{\text{UE}}^{(2)}\right|^2\leqslant\epsilon_1,\label{problem7:constraint1}\\
&\left\|\mathbf{w}_{\text{UE}}^{(2)}\right\|=1,
\end{align}
\end{subequations}
where 
\begin{equation}
    \eta_{i}\triangleq\frac{2|\alpha_{\text{U2R}, i}|^2}{|\alpha_{\text{U2R}, 2}|^2+|\alpha_{\text{U2R}, 3}|^2},~i=2,3.\label{eta_cal}
\end{equation}

The problem (P7) is a non-convex QCQP. Responding to this, we propose two sensing-based beamforming schemes, i.e., the S-SDR and S-MBS beamforming schemes.

1) \textbf{\textcolor{black}{S-SDR beamforming scheme:}}
Note $\left|\mathbf{c}^{\text{H}}\left(u_{\text{U2R}, i}^{\text{D}}\right)\mathbf{w}_{\text{UE}}^{(2)}\right|^{2}=\mathbf[\mathbf{w}_{\text{UE}}^{(2)}]^{\text{H}}\mathbf{P}_{i}\mathbf{w}_{\text{UE}}^{(2)}$, where 
\begin{equation}\mathbf{P}_{i}\triangleq\mathbf{c}\left(u_{\text{U2R}, i}^{\text{D}}\right)\mathbf{c}^{\text{H}}\left(u_{\text{U2R}, i}^{\text{D}}\right),~i=1,2,3,
\label{P_i}
\end{equation}
and the objective function (\ref{problem:7 opt obj}) becomes $\varrho\sum_{i=2,3}\eta_{i}\mathbf[\mathbf{w}_{\text{UE}}^{(2)}]^{\text{H}}\mathbf{P}_{i}\mathbf{w}_{\text{UE}}^{(2)}+(1-\varrho)\mathbf[\mathbf{w}_{\text{UE}}^{(2)}]^{\text{H}}\mathbf{P}_{1}\mathbf{w}_{\text{UE}}^{(2)}$.

Similarly, define 
\begin{equation}
 \mathbf{P}_{\kappa}\triangleq\mathbf{c}\left(u_{\text{U2R},2}^{\text{D}}\right)\mathbf{c}^{\text{H}}\left(u_{\text{U2R},2}^{\text{D}}\right)-\kappa\mathbf{c}\left(u_{\text{U2R},3}^{\text{D}}\right)\mathbf{c}^{\text{H}}\left(u_{\text{U2R},3}^{\text{D}}\right), \label{P_k} 
\end{equation}
and the constraint (\ref{problem7:constraint1}) becomes $0\leqslant\mathbf[\mathbf{w}_{\text{UE}}^{(2)}]^{\text{H}}\mathbf{P}_{\kappa}\mathbf{w}_{\text{UE}}^{(2)}\leqslant\epsilon_1$.

Therefore, we rewrite the problem (P7) as
\begin{subequations}
\begin{align}\label{problem final matrixed}
    \text{(P8)}:~\max _{\mathbf{w}_{\text{UE}}^{(2)}} 
    ~&\varrho\sum_{i=2,3}\eta_{i}\mathbf[\mathbf{w}_{\text{UE}}^{(2)}]^{\text{H}}\mathbf{P}_{i}\mathbf{w}_{\text{UE}}^{(2)}+(1-\varrho)\mathbf[\mathbf{w}_{\text{UE}}^{(2)}]^{\text{H}}\mathbf{P}_{1}\mathbf{w}_{\text{UE}}^{(2)},\\
    \text { s.t. }
    &0\leqslant\mathbf[\mathbf{w}_{\text{UE}}^{(2)}]^{\text{H}}\mathbf{P}_{\kappa}\mathbf{w}_{\text{UE}}^{(2)}\leqslant\epsilon_1,(\ref{constraint:UE}).
\end{align}
\end{subequations}

Noticing $[\mathbf{w}_{\text{UE}}^{(2)}]^{\text{H}}\mathbf{P}_{i}\mathbf{w}_{\text{UE}}^{(2)}=\text{tr}(\mathbf{P}_{i}\mathbf{w}_{\text{UE}}^{(2)}[\mathbf{w}_{\text{UE}}^{(2)}]^{\text{H}})$, we apply the change of variables $\mathbf{W}_{\text{UE}}^{(2)}=\mathbf{w}_{\text{UE}}^{(2)}[\mathbf{w}_{\text{UE}}^{(2)}]^{\text{H}}$, satisfying $\mathbf{W}_{\text{UE}}^{(2)}\succeq \mathbf{0}$ and $\text{rank}(\mathbf{W}_{\text{UE}}^{(2)})=1$, and we convert the problem (P8) into
 \begin{subequations}
\begin{align}\label{problem final traced}
    \text{(P9)}:~\max _{\mathbf{W}_{\text{UE}}^{(2)}} 
    ~&\varrho\sum_{i=2,3}\eta_{i}\text{tr}(\mathbf{P}_{i}\mathbf{W}_{\text{UE}}^{(2)})+(1-\varrho)\text{tr}(\mathbf{P}_{1}\mathbf{W}_{\text{UE}}^{(2)}),\\
    \text { s.t. }
    &0\leqslant\text{tr}(\mathbf{P}_{\kappa}\mathbf{W}_{\text{UE}}^{(2)})\leqslant\epsilon_1,\\ &\text{rank}(\mathbf{W}_{\text{UE}}^{(2)})=1,\label{RankOneCondition}\\
    &\mathbf{W}_{\text{UE}}^{(2)}\succeq\mathbf{0},\left\|\mathbf{W}_{\text{UE}}^{(2)}\right\|^2=1.\label{semidefinite_and_power_W_ue_constraint}
\end{align}
\end{subequations}

To tackle the non-convexity of the rank-one constraint (\ref{RankOneCondition}), we apply the SDR technique to reduce the problem (P9) to
\begin{subequations}
\begin{align}\label{problem final SDP}
    \text{(P10):}~\max _{\mathbf{W}_{\text{UE}}^{(2)}} 
    ~&\varrho\sum_{i=2,3}\eta_{i}\text{tr}(\mathbf{P}_{i}\mathbf{W}_{\text{UE}}^{(2)})+(1-\varrho)\text{tr}(\mathbf{P}_{1}\mathbf{W}_{\text{UE}}^{(2)}),\\
    \text { s.t. }
    &0\leqslant\text{tr}(\mathbf{P}_{\kappa}\mathbf{W}_{\text{UE}}^{(2)})\leqslant\epsilon_1,(\ref{semidefinite_and_power_W_ue_constraint}).
\end{align}
\end{subequations}

Problem (P10) is a standard convex semi-definite program (SDP), which can be optimally solved by convex optimization solvers such as CVX \cite{boyd2004convex}. However, due to the relaxation, solving the relaxed (P10) may not lead to a rank-one solution that satisfies constraint (\ref{RankOneCondition}), which indicates that the optimal objective value of (P10) can only serve as an upper bound of (P7). Nevertheless, many reasonable heuristic methods are proposed to meet the rank-one constraint \cite{luo2010semidefinite}. Here, we adopt the Gaussian randomization (GR) method to meet the rank-one condition. Specifically, we first obtain the eigenvalue decomposition of $\mathbf{W}_{\text{UE}}^{(2)}$ as $\mathbf{W}_{\text{UE}}^{(2)}=\mathbf{U}\mathbf{\Sigma}\mathbf{U}^{\text{H}}$, where $\mathbf{U}=[\boldsymbol{e}_1,\cdots,\boldsymbol{e}_{N_{\text{UE}}}]$ and $\mathbf{\Sigma}=\operatorname{diag}(\lambda_1,\cdots,\lambda_{N_{\text{UE}}})$ are a unitary matrix and a diagonal matrix, respectively. Then, we obtain a suboptimal solution to (P7) as $\bar{\mathbf{w}}_{\text{UE}}^{(2)}=\mathbf{U}\mathbf{\Sigma}^{(1/2)}\mathbf{r}$, where $\mathbf{r}\in\mathbb{C}^{N_\text{UE}\times1}$ is a random vector generated according to $\mathbf{r}\in\mathcal{CN}\left(0,\mathbf{I}_{N_{\text{UE}}}\right)$ with $\mathcal{CN}\left(0,\mathbf{I}_{N_{\text{UE}}}\right)$ denoting the CSCG distribution with zero mean and covariance matrix $\mathbf{I}_{N_{\text{UE}}}$. Let $l_{\text{GR}}$ denote the number of iterations for GR process. We independently generate $l_{\text{GR}}$ Gaussian random vectors $\mathbf{r}$'s and the value of objective function in (P7) is approximated as the maximum one attained by the best $\bar{\mathbf{w}}_{\text{UE}}^{(2)}$ among all $\mathbf{r}$'s. Finally, the feasible solution $\mathbf{w}_{\text{UE}}^{(2)}$ to the problem (P7) can be recovered by $\mathbf{w}_{\text{UE}}^{(2)\star}=\frac{\bar{\mathbf{w}}_{\text{UE}}^{(2)}}{\left\|\bar{\mathbf{w}}_{\text{UE}}^{(2)}\right\|}$. \textcolor{black}{The detailed process of solving the problem (P1) via the S-SDR beamforming scheme is given in Algorithm \ref{alg:SDR}.}

\begin{algorithm}[htp]
\caption{Proposed S-SDR Beamforming Algorithm}
\label{alg:SDR}
\begin{algorithmic}[1]
\REQUIRE $\varrho$, $\mathbf{q}_{\text{BS}}$, $\mathbf{q}_{i}$, $\hat{\mathbf{q}}_{\text{UE}}^{(1)}$, $l_{\text{GR}}$ and the value of the objective function of (P7) $\bar{f}$.
\STATE Compute $\hat{u}_{\text{U2R}, i}^{\text{A},(1)}( i=1,2,3)$ according to (\ref{subsubsection_4}), obtain $\hat{u}_{\text{U2R}, i}^{\text{D},(1)}=-\hat{u}_{\text{U2R}, i}^{\text{A},(1)}$ and compute $\mathbf{c}\left(u_{\text{U2R}, i}^{\text{D}}\right)$ according to (\ref{channel_response_vector_c}).
\STATE Acquire optimal solutions to (P4-a$'$), i.e., $\mathbf{w}_{\text{BS}}^{(2)\star}$ and $\boldsymbol{\xi}^{(2)\star}$, via (\ref{com_result_1}) and (\ref{com_result_2}), respectively.
\STATE Compute $\kappa$, $\epsilon_1$ and $\eta_{i}$,  according to (\ref{kappa_cal_and_epsilon_cal}) and (\ref{eta_cal}).
\STATE Compute $\mathbf{P}_{i}(i=1,2,3)$ and $\mathbf{P}_{\kappa}$ according to (\ref{P_i}) and (\ref{P_k}), respectively.
\STATE Obtain $\mathbf{W}_{\text{UE}}^{(2)}$ by solving problem (P10) with CVX.
\STATE Obtain $\mathbf{U}$ and $\mathbf{\Sigma}$ by applying eigenvalue decomposition to $\mathbf{W}_{\text{UE}}^{(2)}$.
\STATE initialize the value of $\bar{f}$.
\REPEAT 
    \STATE Randomly generate $\mathbf{r}\in\mathcal{CN}\left(0,\mathbf{I}_{N_{\text{UE}}}\right)$ and set $\tilde{\mathbf{w}}_{\text{UE}}^{(2)}=\mathbf{U}\mathbf{\Sigma}^{(1/2)}\mathbf{r}$.
    \STATE Compute the value of the objective function $\tilde{\mathrm{f}}$ corresponding to $\tilde{\mathbf{w}}_{\text{UE}}^{(2)}$ via (\ref{problem:7 opt obj}).
    \IF{$\tilde{\mathrm{f}} \geq \bar{\mathrm{f}}$}\label{alg1:convergence begin}
        \STATE Let $\bar{\mathrm{f}}=\tilde{\mathrm{f}}$ and set $\bar{\mathbf{w}}_{\text{UE}}^{(2)}=\tilde{\mathbf{w}}_{\text{UE}}^{(2)}$. 
    \ENDIF \label{alg1:convergence end}
    \UNTIL The number of iterations reaches $l_{\text{GR}}$.
\STATE Set $\mathbf{w}_{\text{UE}}^{(2)\star}=\frac{\bar{\mathbf{w}}_{\text{UE}}^{(2)}}{\left\|\bar{\mathbf{w}}_{\text{UE}}^{(2)}\right\|}$.
\ENSURE $\mathbf{w}_{\text{BS}}^{(2)\star}$, $\boldsymbol{\xi}^{(2)\star}$ and $\mathbf{w}_{\text{UE}}^{(2)\star}$.
    \end{algorithmic}
\end{algorithm}

2) \textbf{\textcolor{black}{S-MBS beamforming scheme:}}
Although the S-SDR beamforming algorithm can properly solve the problem (P7), it is time-consuming due to the high complexity caused by the use of the SDR technique. To facilitate the practical implementation, \textcolor{black}{we propose a low-complexity S-MBS beamforming algorithm for solving the problem (P7) by maneuvering multiple beams of the UE}. Its main idea is that a suboptimal solution to the problem (P7) can be obtained as the weighted sum of three basic beams respectively steering into the directions of the reflecting sub-surface and the two sensing sub-surfaces. Specifically, the beamforming vector steering to the $i$-th sub-surface is $\mathbf{w}_{i}^{(2)}=\frac{1}{\sqrt{N_{\text{UE}}}}\mathbf{c}\left(u_{\text{U2R}, i}^{\text{D}}\right)$. As such, a suboptimal solution to the problem (P7) can be expressed in the following form
\begin{equation}
\mathbf{w}_{\text{UE}}^{(2)}=\frac{\sum_{i=1,2,3}\psi_{i}\mathbf{w}_{i}^{(2)}}{\left\|\sum_{i=1,2,3}\psi_{i}\mathbf{w}_{i}^{(2)}\right\|},\label{weighted coefficients of PSO}
\end{equation}
where $\psi_i\in[0,1]~(i=1,2,3)$ is the weight factor. Let $\boldsymbol{\psi}=[\psi_{1},\psi_{2},\psi_{3}]^{\text{T}}$. Then, by substituting (\ref{weighted coefficients of PSO}) into the problem (P7), we optimize the weight vector $\boldsymbol{\psi}$ instead of the UE beamforming vector $\mathbf{w}_{\text{UE}}^{(2)}$, and transform the problem (P7) into
\begin{subequations}
\begin{align}
    \text{(P11)}:~\max _{\boldsymbol{\psi}} 
    ~&\varrho\sum_{i=2,3} \eta_{i}F_i+(1-\varrho)F_1,\\
    \text { s.t. }
    &|F_2-\kappa F_3|\leqslant\epsilon_1,\label{PSO_constraint_to_penaly}\\
    &\left\|\mathbf{w}_{\text{UE}}^{(2)}\right\|=1,
\end{align}
\end{subequations}
where $F_i\triangleq\left|\mathbf{c}^{\text{H}}(u_{\text{U2R},i}^{\text{D}})\mathbf{w}_{\text{UE}}^{(2)}\right|^2,~i=1,2,3$. To solve (P11), we incorporate the constraint (\ref{PSO_constraint_to_penaly}) by exploiting the penalty function method with the penalty function $P(x)=-\mu x$, where $\mu>0$ is the penalty parameter which is used to scale the penalty function and $x$ is the variable. As such, we reformulate the problem (P11) as
\begin{subequations}
\begin{align}
    \text{(P12)}:~\max _{\boldsymbol{\psi}} 
    ~&\varrho\sum_{i=2,3} \eta_{i}F_i+(1-\varrho)F_1-\mu\left(\left|F_2-\kappa F_3\right|-\epsilon_1 \right),\label{PSO objective function}\\
    \text { s.t. }
    &\left\|\mathbf{w}_{\text{UE}}^{(2)}\right\|=1.
\end{align}
\end{subequations}

In the above objective function, since the penalty part corresponds to the sensing performance weighted by  $\varrho$, we deliberately set the penalty parameter as  $\mu=2\varrho$ such that the sensing performance, which is determined by the first and third terms, improves with the increase of the trade-off factor $\varrho$. \textcolor{black}{To solve (P12), we adopt the method of particle swarm optimization (PSO)}. First, we generate $N_\text{p}$ particles, and each particle has two properties, i.e., the position and the velocity, which respectively represent a possible vector of the weight factor, i.e. $\boldsymbol\psi$, in (75), and the corresponding search direction. For the $k$-th particle in the $l$-th iteration of the algorithm, we define its position and velocity as $\boldsymbol{\psi}_{k,l}$ and $\boldsymbol{v}_{k,l}$, respectively, where $[\boldsymbol{\psi}_{k,l}]_{j}\in[0,1]$ and $[\boldsymbol{v}_{k,l}]_{j}\in[v_{\text{min}},v_{\text{max}}]$ $(j=1,2,3)$. Let the value of (\ref{PSO objective function}) be the fitness value in the PSO process and denote it as $fitness(\boldsymbol{\psi}_{k,l})$ for given $\boldsymbol{\psi}_{k,l}$. We evaluate the corresponding fitness value for each particle in each iteration, and record the best fitted position $\tilde{\boldsymbol{\psi}}_{k}$ for the $k$-th particle and the globally best fitted position $\boldsymbol{\psi}_{\text{Best}}$ for all particles. 
For the $l$-th iteration, the $k$-th particle updates its velocity $\boldsymbol{v}_{k,l}$ via
\begin{equation}
\label{velocity update}\boldsymbol{v}_{k,l}=c_1\boldsymbol{v}_{k,l-1}+c_2\chi({\boldsymbol{\psi}}_{\text{Best}}-\boldsymbol{\psi}_{k,l-1})+c_3\chi(\tilde{\boldsymbol{\psi}}_{k}-\boldsymbol{\psi}_{k,l-1}),
\end{equation}
where $c_i~(i=1,2,3)$ is the weight factor for each term, which represents the learning rate in the PSO process, and $\chi$ is the random values which is generated via $\chi\sim U(0,1)$.

Noticing that $v_{\text{min}}\leqslant[\boldsymbol{v}_{k,l}]_{j}\leqslant v_{\text{max}}$ $(j=1,2,3)$, we have $[\boldsymbol{v}_{k,l}]_{j}=\operatorname{max}(v_{\text{min}},[\boldsymbol{v}_{k,l}]_{j}),~[\boldsymbol{v}_{k,l}]_{j}=\operatorname{min}(v_{\text{max}},[\boldsymbol{v}_{k,l}]_{j})$, and then we update the position $\boldsymbol{\psi}_{k,l}$ via 
\begin{equation}
\label{position update}\boldsymbol{\psi}_{k,l}=\boldsymbol{\psi}_{k,l-1}+\boldsymbol{v}_{k,l},
\end{equation}
where we deliberately limit every element of $\boldsymbol{\psi}_{k,l}$ in $[0,1]$. Moreover, $\boldsymbol{\psi}_{k,0}$ and $\boldsymbol{v}_{k,0}$ respectively represent the initial position and velocity of the $k$-th particle. In the initialization, the
velocity and position of each particle are set randomly within the search space. \textcolor{black}{The detailed process of solving the problem (P1) via the S-MBS beamforming scheme is summarized in Algorithm \ref{alg:PSO}.}

\begin{algorithm}
    \caption{\textcolor{black}{Proposed S-MBS Beamforming Algorithm}}
    \label{alg:PSO}
    \begin{algorithmic}[1]
        \REQUIRE $\varrho$, $\mathbf{q}_{\text{BS}}$, $\mathbf{q}_{i}$, $\hat{\mathbf{q}}_{\text{UE}}^{(1)}$, the number of particles $N_\text{p}$ and the number of iterations $l_{\text{p}}$.
        \STATE Compute $\hat{u}_{\text{U2R}, i}^{\text{A},(1)}( i=1,2,3)$ according to (\ref{subsubsection_4}), obtain $\hat{u}_{\text{U2R}, i}^{\text{D},(1)}=-\hat{u}_{\text{U2R}, i}^{\text{A},(1)}$ and compute $\mathbf{c}(u_{\text{U2R}, i}^{\text{D}})$ according to (\ref{channel_response_vector_c}).
        \STATE Acquire optimal solutions to (P4-a$'$), i.e., $\mathbf{w}_{\text{BS}}^{(2)\star}$ and $\boldsymbol{\xi}^{(2)\star}$, via (\ref{com_result_1}) and (\ref{com_result_2}), respectively.
        \STATE Compute $\kappa$ and $\eta_{i}$ according to (\ref{kappa_cal_and_epsilon_cal}) and (\ref{eta_cal}), and let $\mu=2\varrho$.
        \FOR{$k$-th particle}
        \STATE Initialize the velocity $\boldsymbol{v}_{k,0}$ and the position $\boldsymbol{\psi}_{k,0}$ randomly.
        \STATE Evaluate the fitness via (\ref{PSO objective function}) and set $\tilde{\boldsymbol{\psi}}_{k}=\boldsymbol{\psi}_{k,0}$.
        \ENDFOR
        \STATE Set ${\boldsymbol{\psi}}_{\text{Best}}=\tilde{\boldsymbol{\psi}}_{1}$ as the initial value.
        \FOR{$l$-th iteration}
        \FOR{$k$-th particle}
        \STATE Update the velocity $\boldsymbol{v}_{k,l}$ and the position $\boldsymbol{\psi}_{k,l}$  via (\ref{velocity update})-(\ref{position update}).
        \STATE Evaluate the fitness via (\ref{PSO objective function}).
        \IF {$fitness(\boldsymbol{\psi}_{k,l})> fitness(\tilde{\boldsymbol{\psi}}_{k})$} \label{alg2:convergence begin}
        \STATE $\tilde{\boldsymbol{\psi}}_{k}=\boldsymbol{\psi}_{k,l}$.
        \ENDIF
        \IF {$fitness(\tilde{\boldsymbol{\psi}}_{k})> fitness({\boldsymbol{\psi}}_{\text{Best}})$}
        \STATE ${\boldsymbol{\psi}}_{\text{Best}}=\tilde{\boldsymbol{\psi}}_{k}$.
        \ENDIF          \label{alg2:convergence end}
        \ENDFOR
        \ENDFOR
        
        \STATE Set $\mathbf{w}_{\text{UE}}^{(2)\star}=\frac{\bar{\mathbf{w}}_{\text{UE}}^{(2)}}{\left\|\bar{\mathbf{w}}_{\text{UE}}^{(2)}\right\|}$, where $\bar{\mathbf{w}}_{\text{UE}}^{(2)}=\sum_{j=1,2,3}[{\boldsymbol{\psi}}_{\text{Best}}]_j\mathbf{w}_{i}^{(2)}$.
        \ENSURE $\mathbf{w}_{\text{BS}}^{(2)\star}$, $\boldsymbol{\xi}^{(2)\star}$ and $\mathbf{w}_{\text{UE}}^{(2)\star}$.
    \end{algorithmic}
\end{algorithm}

Additionally, the complexity and convergence analysis of the proposed two sensing-based beamforming schemes in phase 2 are provided as follows.

\textbf{Complexity analysis:} The common part of the proposed two beamforming schemes in phase 2 deals with the problem of optimizing the active beamforming vector of the BS $\mathbf{w}_{\text{BS}}^{(2)}$ and the passive beamforming vector of the reflecting sub-surface $\boldsymbol{\xi}^{(2)}$. The optimal beamforming vectors $\mathbf{w}_{\text{BS}}^{(2)\star}$ and $\boldsymbol{\xi}^{(2)\star}$ are acquired via (\ref{com_result_1}) and (\ref{com_result_2}), respectively. Hence, the complexity of the common part is mainly determined by calculating (\ref{com_result_1}) and (\ref{com_result_2}), and the corresponding complexity is $\mathcal{O}(N_{\text{BS}}+M_{1})$. Additionally, in order to solve the non-convex QCQP problem (P7), Algorithm \ref{alg:SDR} adopts the SDR method and solves the follow-up SDP problem (P10) via CVX. By using the interior point method, the worst-case complexity of solving the SDP problem (P10) is $\mathcal{O}(N_{\text{UE}}^{4.5}\text{log}(\frac{1}{\iota}))$, where $\iota > 0$ is the solution accuracy\cite{luo2010semidefinite,SDP_Complexity_TIT_mobasher2007near}. The subsequent GR procedure is applied to meet the rank-one constraint, the complexity of which is $\mathcal{O}(l_{\text{GR}}N_{\text{UE}})$. Therefore, the complexity of Algorithm \ref{alg:SDR} is $\mathcal{O}(N_{\text{BS}}+M_{1}+N_{\text{UE}}^{4.5}\text{log}(\frac{1}{\iota})+l_{\text{GR}}N_{\text{UE}})$. While for Algorithm \ref{alg:PSO}, due to the transformation in (\ref{weighted coefficients of PSO}), the complexity of Algorithm \ref{alg:PSO} is mainly determined by the computation of (\ref{PSO objective function}) in each iteration, and thus reduced to $\mathcal{O}(N_{\text{BS}}+M_{1}+N_{\text{p}}l_{\text{p}}N_{\text{UE}})$.

\textbf{Convergence analysis:} For Algorithm \ref{alg:SDR}, its convergence is guaranteed since the value of the objective function of (P7) in each iteration of the GR procedure is updated non-decreasingly and the optimal value is bounded from above owing to the SNR constraint. Similarly, for Algorithm \ref{alg:PSO}, the value of its objective function, i.e. the optimal fitness of all particles, is updated increasingly over iterations and the SNR constraint bounds its value from above.

\vspace{-3mm} \section{Simulation Results}
\begin{figure}[htbp]
  \begin{minipage}[t]{0.5\linewidth}
    \centering
    \includegraphics[width=0.9\linewidth]{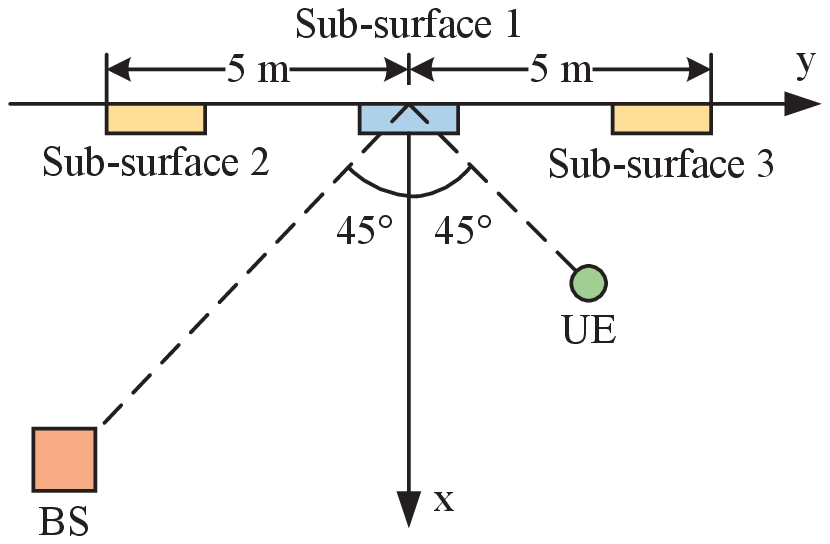}
    \caption{Simulation setup (top view).}
    \label{sim_setup_common}
  \end{minipage}%
  \begin{minipage}[t]{0.5\linewidth}
    \centering
    \includegraphics[width=1\linewidth]{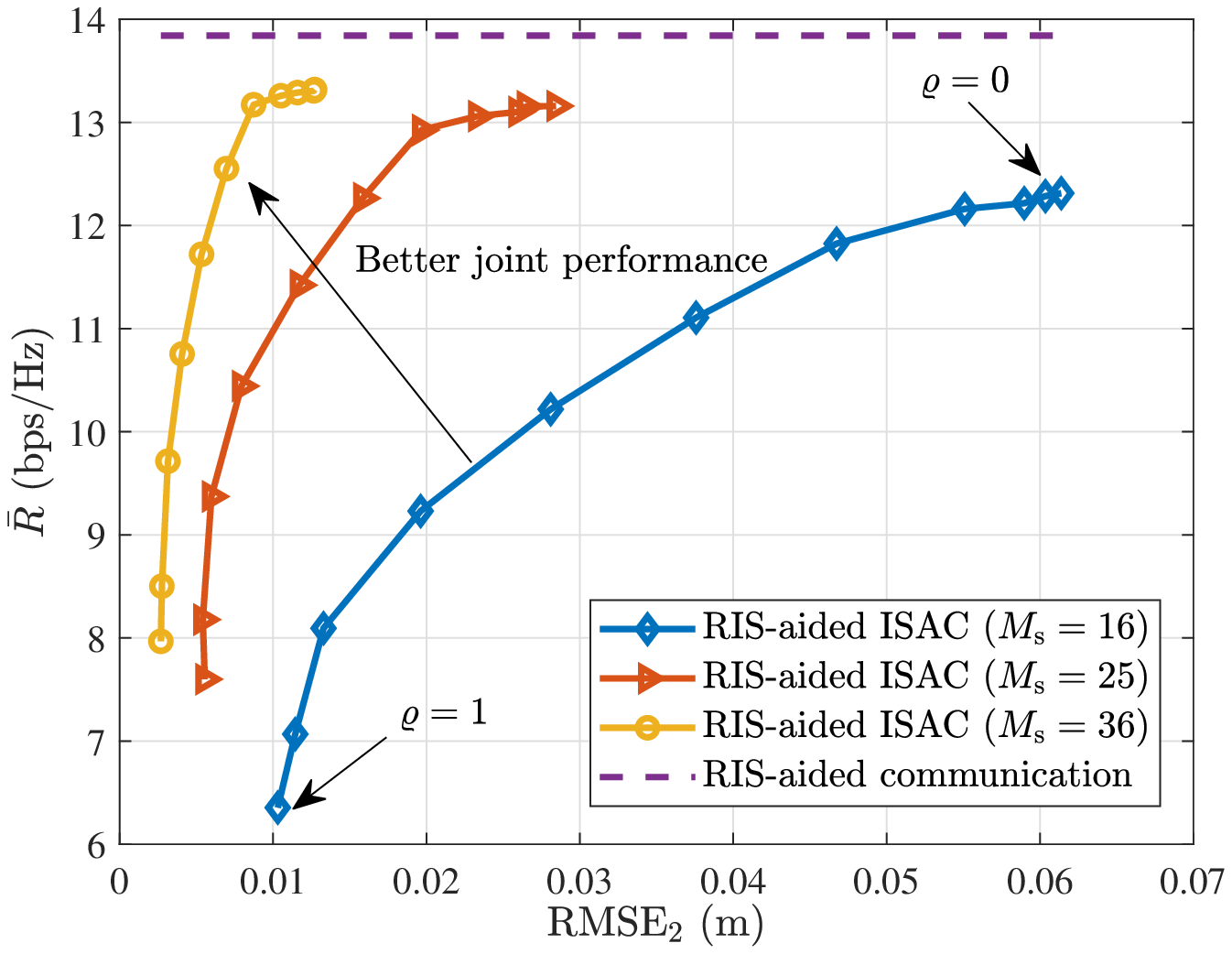}
    \caption{\color{black}Impact of the sensing elements on the proposed RIS-aided ISAC system.}
    \label{sim_A_1}
  \end{minipage}
\end{figure}

In this section, we provide simulation results to demonstrate the performance of the proposed RIS-aided ISAC system and the sensing-based beamforming schemes. The simulation setup is shown in Fig. \ref{sim_setup_common}, where the UE is on the horizontal floor, the RIS is 3 meters (m) above the horizontal floor, and the BS is 20 m above the horizontal floor. The distances from the UE and the BS to the first sub-surface are set to be $d_{\text{U2R},1}=5$ m and $d_{\text{B2R},1}=50$ m, respectively. The path loss exponents from the RIS to the BS, from the UE to the RIS, and from one sub-surface to other sub-surfaces are set as $2.3$, $2.2$ and $2.1$, respectively. The path loss at the reference distance of $1 \text{~m}$ is set as $30$ dB. Unless otherwise specified, the following setup is used: $N_{\text{BS}}=16$, $M_{1}=20 \times 20$, $M_{2}=M_{3}=M_{\text{s}}=6 \times 6$, $\tau_{1}=5$, $\tau_{2}=95$, $T=100$, $\rho=20 \text{~dBm}$, noise power $\sigma_{0}^{2}=-80 \text{~dBm}$, simulation setup in Fig. \ref{sim_setup_common} and the proposed
S-SDR beamforming algorithm are adopted. Besides, the root mean square error (RMSE) is adopted to measure the performance of location sensing. Let $\text{RMSE}_{n}$ represent the RMSE in phase $n$ and define it as
$
\text{RMSE}_{n}=\sqrt{\hat{\mathbf{q}}_{\text{U}}^{(n)}-{\mathbf{q}}_{\text{U}}},~n=1,2.
$

\subsection{Proposed RIS-aided ISAC System vs. RIS-aided Communication System}\label{simulation_subsection_A}
In this subsection, we focus on the comparison of the proposed RIS-aided ISAC system and the RIS-aided communication system. In the RIS-aided communication system, the alternating optimization (AO) scheme assuming perfect CSI \cite{wu2019intelligent} is adopted. We use the average communication rate over the whole coherence block $\bar{R}$ as the metric for measuring the communication performance, and define it as $\bar{R}=\frac{1}{T}\sum_{t=1}^T{R\left( t \right)}$. Moreover, to characterize the JC\&S performance of the RIS-aided ISAC system, we adopt the rate-distortion metric, i.e., $\bar{R}-\text{RMSE}_2$, which indicates the average communication rate achieved under a given RMSE in phase 2.

{\color{black}Fig. \ref{sim_A_1} compares the performance of the proposed RIS-aided ISAC system and the RIS-aided communication system under different numbers of sensing elements. The $\bar R-\text{RMSE}_2$ curves represent the JC\&S performance of the proposed system, as the trade-off factor $\varrho$ ranges from 0 (i.e., communication SNR maximization) to 1 (i.e., sensing SNR maximization). First, we can see that the communication performance of the proposed RIS-aided ISAC system is worse than that of the RIS-aided communication system, since the latter one assumes the perfect CSI is known and spends its all spatial resources on communication. However, despite worse communication performance, the RIS-aided ISAC system realizes the new function of location sensing, and can achieve comparable communication performance to the RIS-aided communication system by adjusting the trade-off factor $\varrho$. Moreover, their communication performance gap decreases with the number of sensing elements $M_\text{s}$, because more accurate sensing information is obtained and used for beamforming design in the RIS-aided ISAC system. Additionally, we can observe that the proposed beamforming scheme can flexibly balance the communication and sensing performance by adjusting the trade-off factor $\varrho$. By sacrificing the performance of communication, the performance of location sensing can be improved and vice versa. For instance, when $M_{s}=16$, the $\text{RMSE}_2$ ameliorates from 0.06 m to 0.01 m as the communication rate decreases from 12.3 bps/Hz to 6.4 bps/Hz.}
    
\begin{figure}[htbp]
\centering
\subfigure[$\bar{R}$ vs. $\tau_1/T$ with different $M_{\text{s}}$ ($\varrho=0,~\rho=10$ dBm,  $d_{\text{U2R},1}=10$ m).]
{
\includegraphics[width=0.47\linewidth]{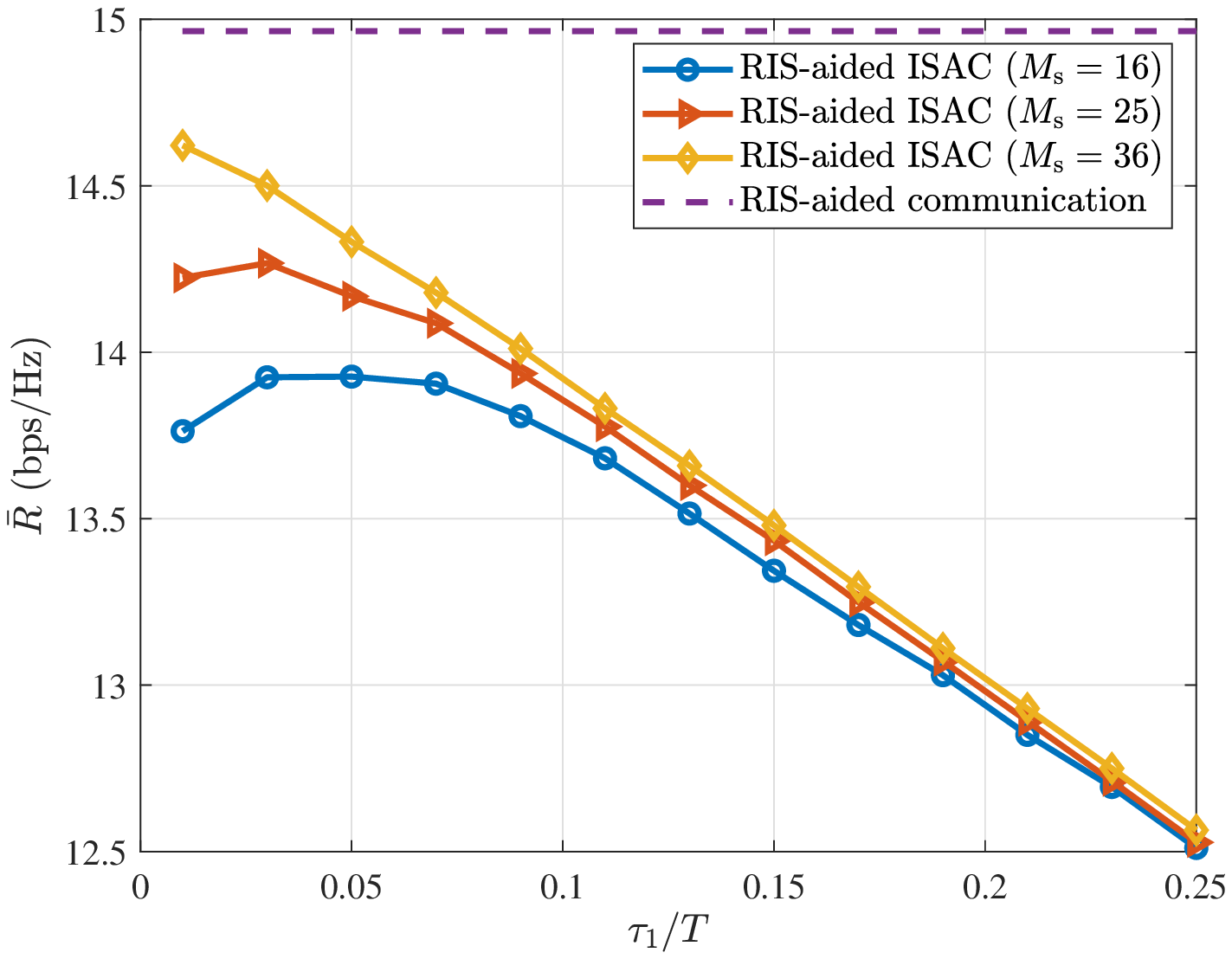}
} 
\subfigure[$\bar{R}$ vs. $\text{RMSE}_2$ with different $\tau_1/T$ ($\rho=10$ dBm).]{\includegraphics[width=0.47\linewidth]{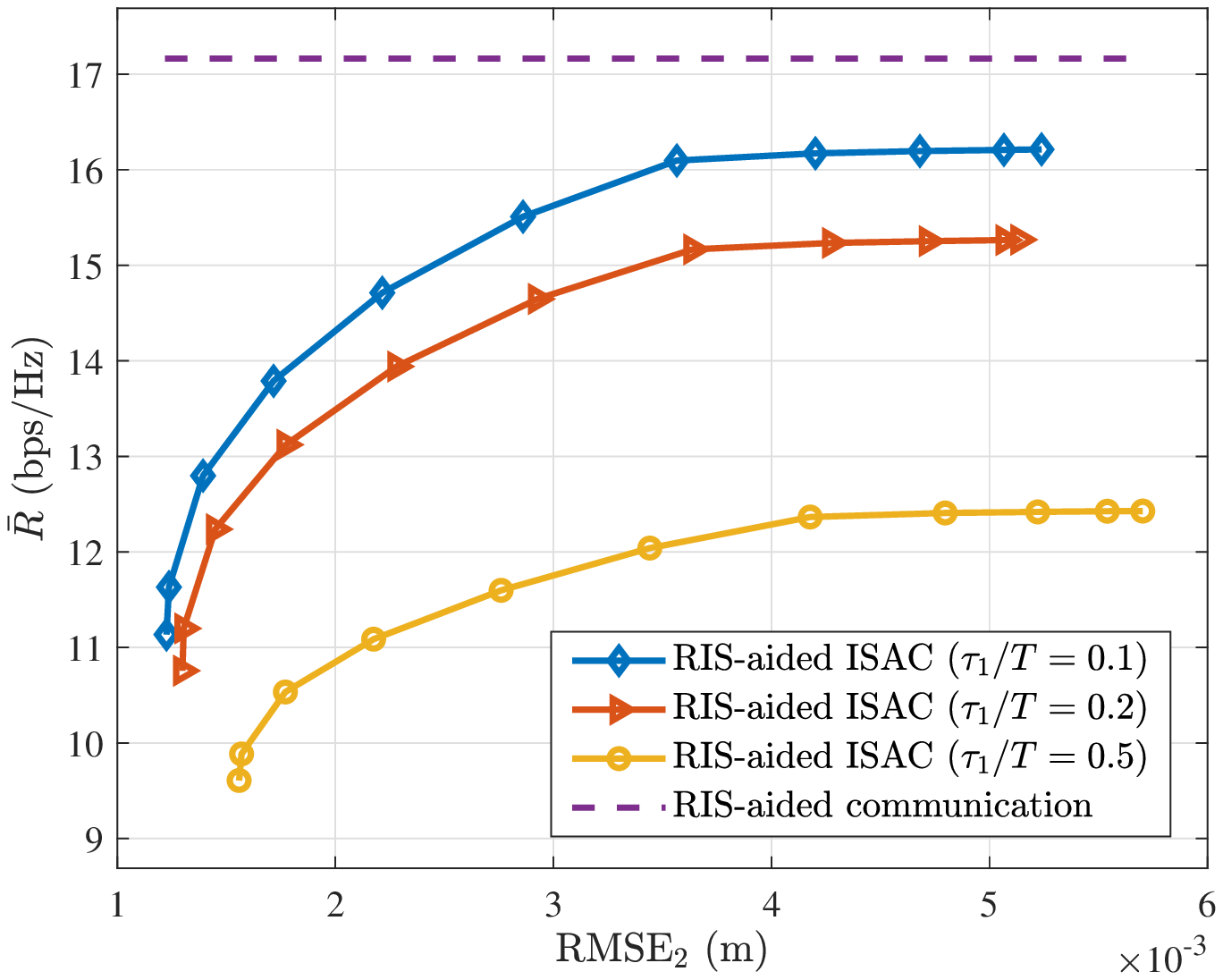} }
\caption{Impact of the phase 1 ratio on the proposed RIS-aided ISAC system.}
\label{sim_A_2}
\end{figure}

Fig. \ref{sim_A_2} compares the performance of the proposed RIS-aided ISAC system and the RIS-aided communication system when different time ratios of phase 1 in the RIS-aided ISAC system are considered. In Fig. 6(a), for all configurations of $\tau_1/T$ and $M_\text{s}$, the average communication rate of the proposed RIS-aided system is lower than that of the RIS-aided communication system. This communication performance gap can be narrowed by increasing the number of sensing elements or properly designing the time ratio $\tau_1/T$. 
In addition, there exists an optimal ratio that maximizes the average communication rate. For example, with $M_{\text{s}}=16$, $\bar R$ increases to its highest point when $\tau_1/T$ changes from 0.01 to 0.05 and then drops when $\tau_1/T$ keeps increasing. This is because when the number of sensing elements $M_{\text{s}}$ and the ratio of phase 1 $\tau_1/T$ are both small, the resulting low positioning accuracy in phase 1 would affect the effectiveness of the sensing-based beamforming design in phase 2, thereby causing the drop of the average communication rate. However, by appropriately increasing the ratio of phase 1, the error of the location sensing can be ameliorated effectively and $\bar R$ rises correspondingly. Moreover, as the number of sensing elements $M_{\text{s}}$ increases, the optimal time ratio becomes smaller. For instance, by increasing $M_\text{s}$ from $16$ to $36$, the best time ratio decreases from 0.05 to 0.01 for the following reasons. On the one hand, when $M_\text{s}$ is large, the sensing accuracy is sufficiently high for effective beamforming design even with small $\tau_1/T$, and excessive improvement of sensing accuracy can only bring marginal benefits to the communication performance. On the other hand, increasing $\tau_1/T$ causes fewer time slots allocated for the high-rate communication in phase 2. More specifically, in Fig. \ref{sim_A_2}(b), the number of sensing elements $M_\text{s}$ is set to be fixed and we focus on the comparison between two systems with different $\tau_1/T$. The ratio of phase 1 affects not only the average communication rate but also the location sensing accuracy. This is expected since, with the aid of the estimated UE's location in phase 1, beamforming in phase 2 is able to effectively improve both communication and sensing performances, and such improvement depends on the value of $\varrho$. Additionally, the results of communication performance in Fig. \ref{sim_A_2}(b) are consistent with the results in Fig. \ref{sim_A_2}(a), where the average communication rate $\bar R$ drops with the increase of $\tau_1/T$.

\begin{figure}
  \begin{minipage}[t]{0.5\linewidth}
    \centering
    \includegraphics[width=1\linewidth]{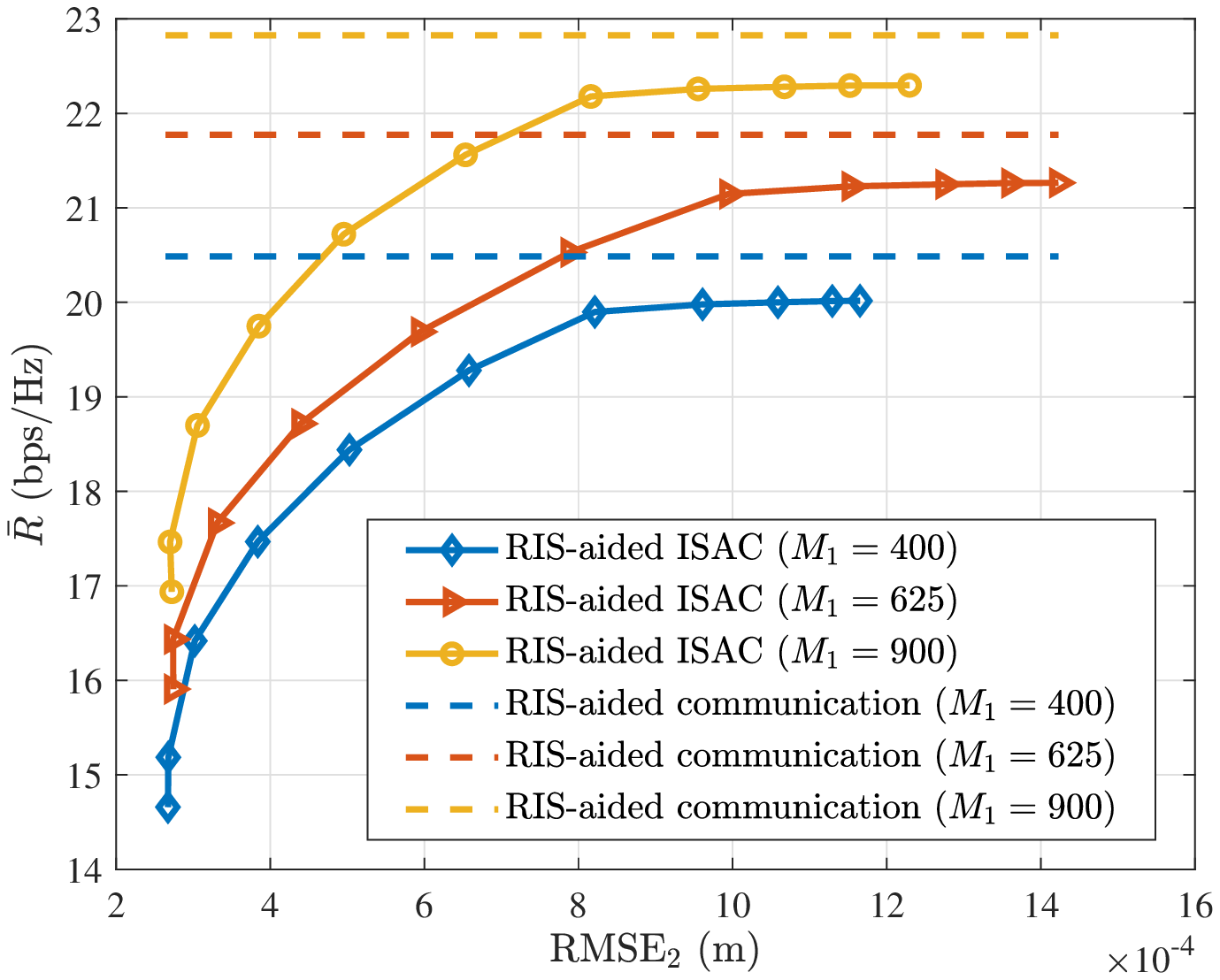}
    \caption{Impact of the number of reflecting elements on the RIS-aided ISAC system.}
    \label{sim_A_3}
  \end{minipage}%
  \begin{minipage}[t]{0.5\linewidth}
    \centering
    \includegraphics[width=1\linewidth]{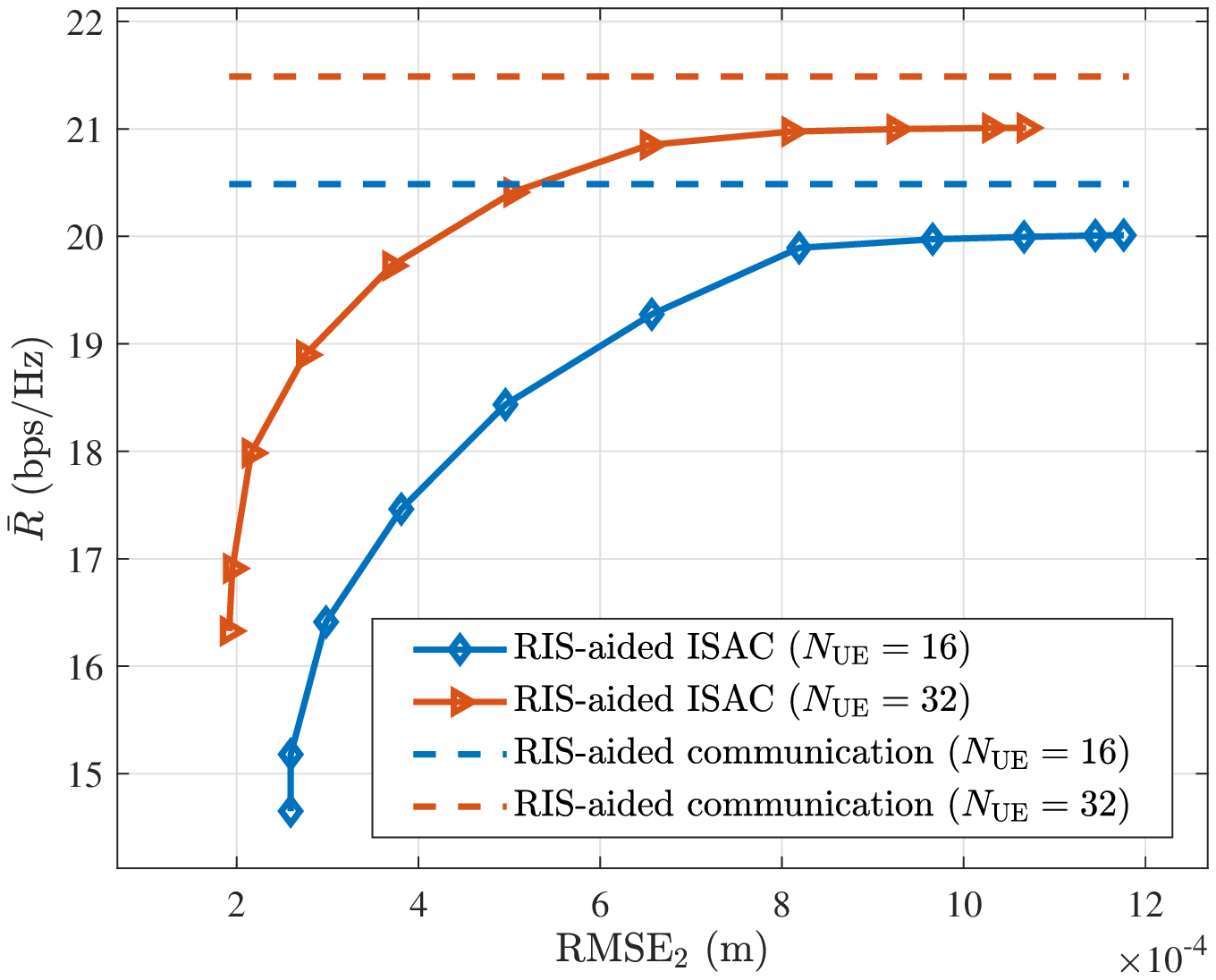}
    \caption{Impact of the number of UE's antennas on the RIS-aided ISAC system.}
    \label{sim_A_4}
  \end{minipage}
\end{figure}

In Fig. \ref{sim_A_3}, we compare the performances of the proposed RIS-aided ISAC system and the RIS-aided communication system with different numbers of reflecting elements. The average communication rate between the two systems is almost unchanged with different $M_1$. However, the performance of the RIS-aided communication system relies on perfect CSI, whose channel estimation overhead increases linearly with $M_1$. \textcolor{black}{By contrast, the channel estimation overhead of the proposed RIS-aided ISAC system is irrelevant with $M_1$, avoiding high overhead of cascaded channel estimation.} In addition, the joint performance improves when the number of passive elements increases. Such improvement is due to the additional spatial resources provided by the RIS with more reflecting elements. 

In Fig. \ref{sim_A_4}, we compare the performance of the proposed RIS-aided ISAC system and the RIS-aided communication system when different numbers of UE's antennas are considered. As the number of UE's antennas becomes larger, the communication performance of both systems improves due to higher beamforming gain, and their gap still remains approximately unchanged. Additionally, increasing the number of UE's antennas significantly improves the joint communication and localization performance of the IRS-aided ISAC system, due to the increased spatial resources provided by more UE's antennas for both communication and location sensing.

\vspace{-3mm} \subsection{How Does Beamforming Affect the Communication-Sensing Trade-off?}\label{simulation_subsection_B}

In this subsection, we focus on answering the question how beamforming affects the trade-off in the RIS-aided ISAC system by comparing the proposed two sensing-based beamforming schemes. \textcolor{black}{To characterize the JC\&S performance, we also adopt $\bar{R}_2-\text{RMSE}_2$ as the metric in this subsection, where $\bar{R}_2$ is the average communication rate in phase 2 and is defined as} 
$
\bar{R}_2=\frac{1}{\tau_2}\sum_{t=\tau _1+1}^T{R\left( t \right)}.
$

\begin{figure}[htbp] 
\centering
\subfigure[$\varrho=0$.]{\includegraphics[width=0.3\textwidth]{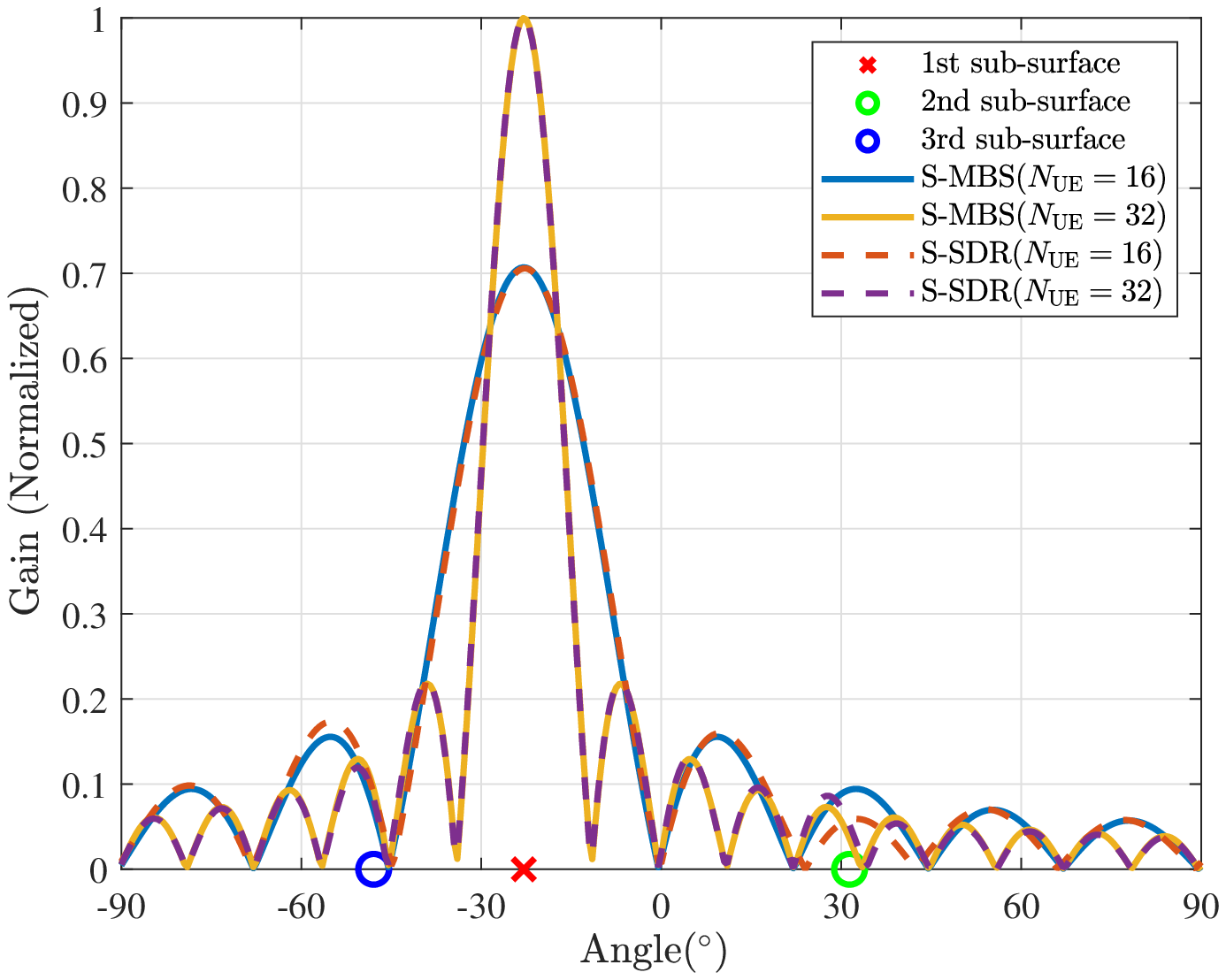} }
\subfigure[$\varrho=0.5$.]{\includegraphics[width=0.3\textwidth]{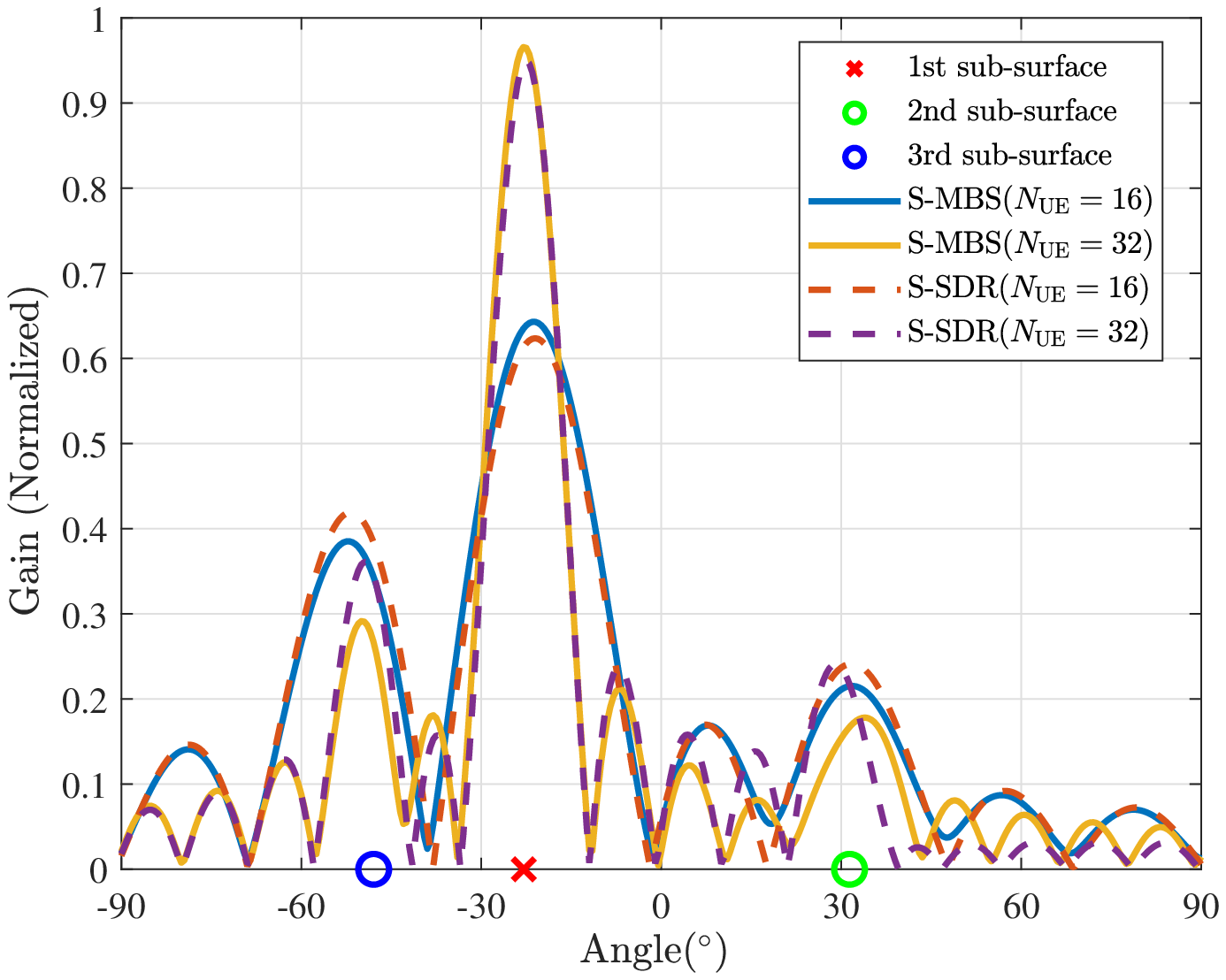} }
\subfigure[$\varrho=1$.]{\includegraphics[width=0.3\textwidth]{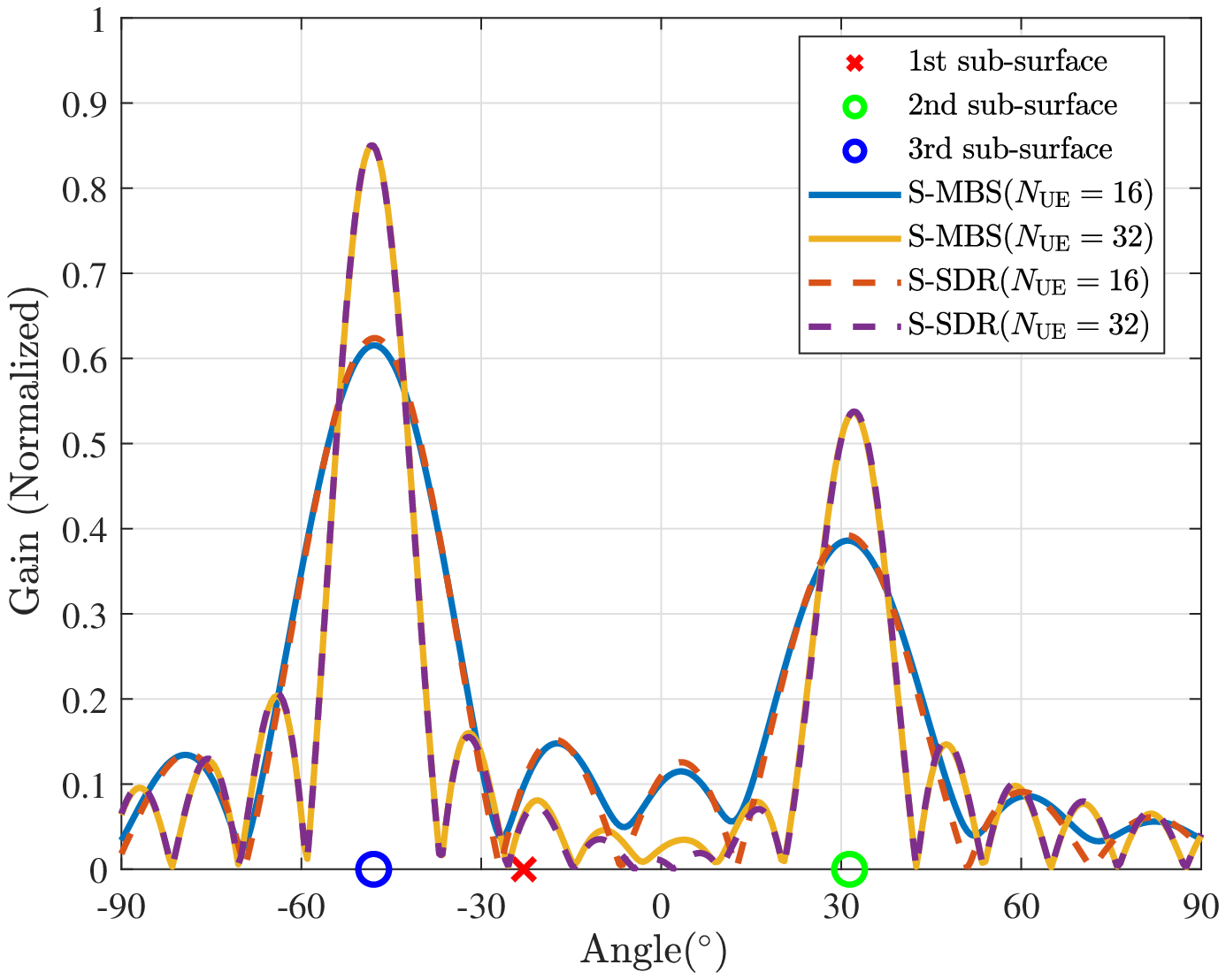} }
\caption{UE's beampattern gains achieved by the proposed sensing-based beamforming scheme.}

\label{sim_B_1}
\vspace{-5mm}
\end{figure}
\hspace*{\fill} 

Fig. \ref{sim_B_1} compares UE's beampattern gains achieved by the proposed two sensing-based beamforming schemes with different $\varrho$, where the location of the UE is set to be (3.46 m, -2.00 m, 0.00 m).  By adjusting the trade-off factor $\varrho$,  desired beam pattern can be generated by the proposed sensing-based beamforming schemes. For instance, in Fig. \ref{sim_B_1}(a), with $\varrho=0$ which corresponds to communication SNR maximization, the main beam is steered towards the reflecting sub-surface for assisting communication, while  in Fig. \ref{sim_B_1}(c), with $\varrho=1$ which corresponds to sensing SNR maximization, the strongest two beams are respectively steered towards the two sensing sub-surface for enhancing location sensing. In Fig. \ref{sim_B_1}(b), with $\varrho=0.5$ which means that both communication and location sensing performances are taken into account, the strongest three beams are respectively steered towards the reflecting sub-surface and the two sensing sub-surface such that both performances can be improved. Moreover, due to the consideration of avoiding the ``cask effect" in estimating AOAs of both sensing sub-surfaces, in Fig. \ref{sim_B_1} (b) and Fig. \ref{sim_B_1} (c), we can readily observe that the beam to the direction of the second sub-surface (the one which is far from the UE) is stronger than the third sub-surface. This helps further improve the location sensing accuracy via UE's beamforming.  Also, increasing the number of UE's antennas improves JC\&S performance due to the higher beamforming gain.

\begin{figure}[htbp]
\centering
\includegraphics[width=0.45\linewidth]{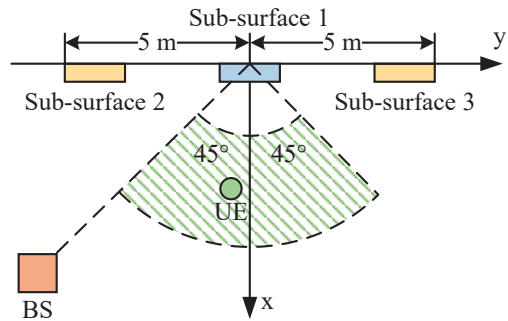} 
\caption{Simulation setup for Fig. \ref{sim_B_2} and Fig. \ref{Response-D-1} (top view).}
\label{sim_setup_B_random}
\end{figure}

\begin{figure}
\centering
\subfigure[$\bar{R}_2$ vs. $\text{RMSE}_2$ with different $\rho$ ($\tau_{1}=20,~T=40$).]{\includegraphics[width=0.47\linewidth]{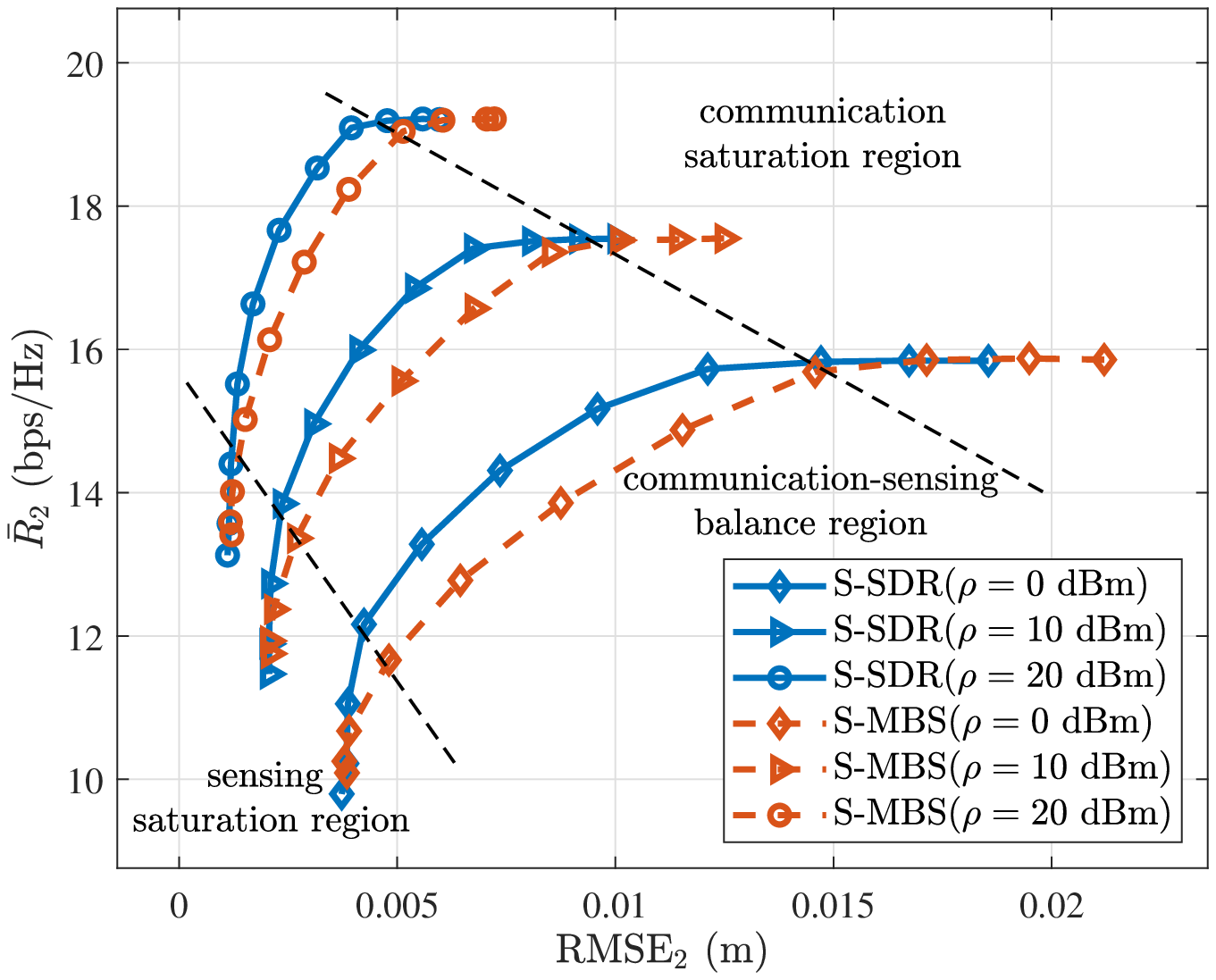}} 
\subfigure[$\bar{R}_2$ vs. $\text{RMSE}_2$ with different $N_{\text{UE}}$ ($\tau_{1}=20,~T=40$).]{\includegraphics[width=0.47\linewidth]{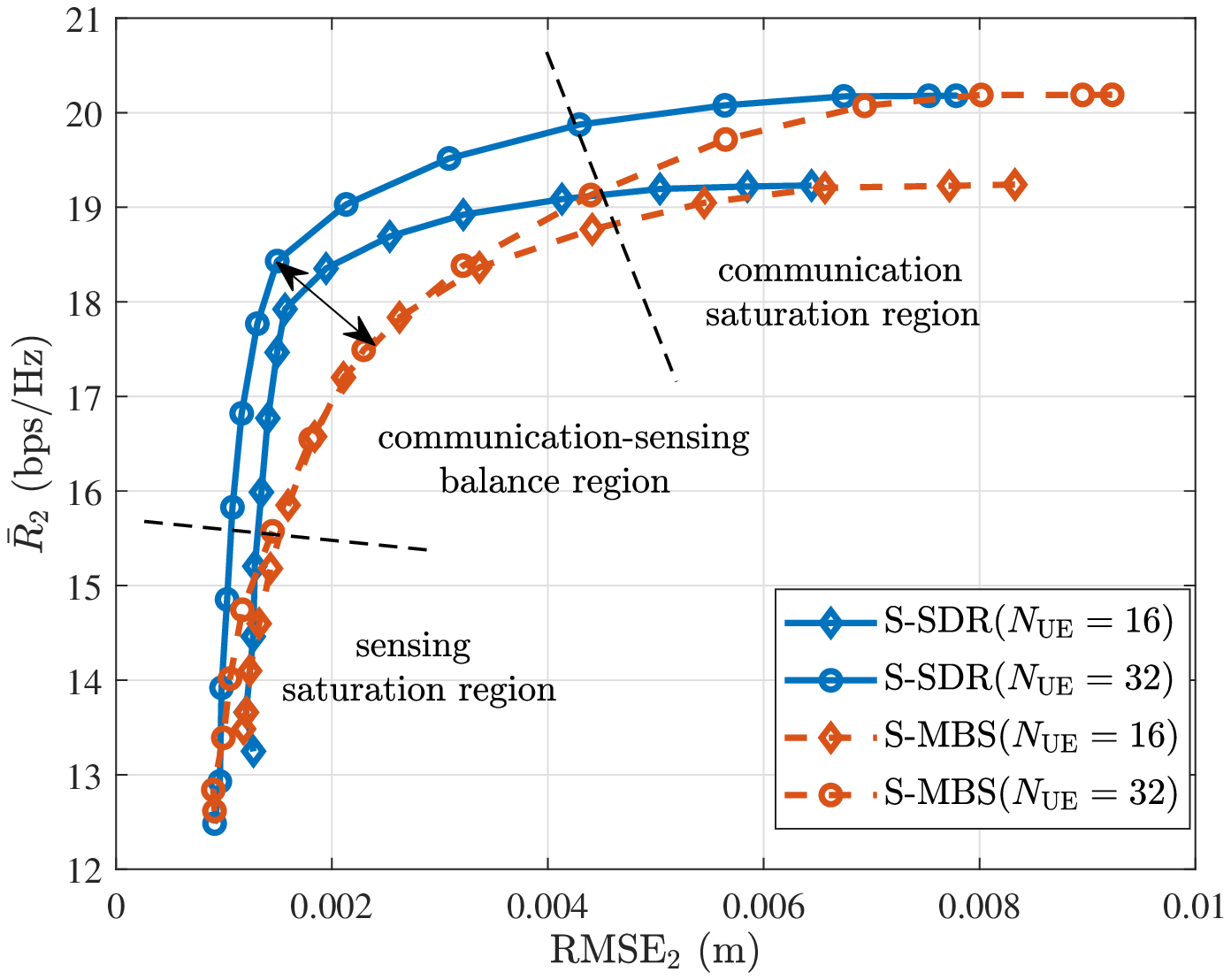} }
\caption{Comparison of the joint performance between proposed beamforming schemes.}
\label{sim_B_2}
\end{figure} 

\textcolor{black}{In Fig. \ref{sim_B_2}, we present the JC\&S performance of the proposed two beamforming schemes with the simulation setup illustrated in Fig. \ref{sim_setup_B_random}, where the UE is randomly distributed in the green area in front of the RIS. Specifically, the 3D-distance from the UE to the first sub-surface $d_{\text{U2R},1}$ is limited in $[5, 10]$ m and the corresponding angle ranges in $[-45, 45]$ degrees.} The rate-RMSE curves can be divided into three regions, i.e., the communication-sensing balance region, communication saturation region and sensing saturation region. In the balance region, the performance of communication and location is well balanced. The performance of one can be effectively improved by sacrificing the performance of another. In the communication/sensing saturation region, even if the sensing/communication performance is sacrificed a lot, little improvement of the communication/sensing performance can be achieved. Generally, we hope that the ISAC system operates in the balance region such that better joint performance can be obtained. Specifically, Fig. \ref{sim_B_2}(a) compares performances of the two proposed beamforming schemes under different transmission power $\rho$. The JC\&S performance of the S-SDR beamforming scheme outperforms that of the S-MBS beamforming scheme, especially in the communication-sensing balance region where both performances are considered to be important. However, their performance gap gradually narrows to triviality in both communication and sensing saturation regions. Also, Fig. \ref{sim_B_2}(b) shows the impact of the number of UE's antennas $N_{\text{UE}}$ on the communication-sensing trade-off. In the balance region, the JC\&S performance of the S-SDR beamforming scheme is superior to that of the S-MBS beamforming scheme, and their performance gap becomes more pronounced with the increase of $N_{\text{UE}}$. While such performance gap is trivial in both communication and sensing saturation regions for any configuration of $N_{\text{UE}}$. This indicates that the S-SDR beamforming scheme has a better ability to balance communication and sensing. Moreover, as has been analysed in Section \ref{section_beamforming}, the complexity gap between the S-SDR beamforming scheme and the S-MBS beamforming scheme enlarges with the increase of $N_{\text{UE}}$, which indicates that the S-SDR beamforming scheme can achieve better joint performance in the balance region with the price of higher computational complexity, whereas the S-MBS beamforming scheme can achieve almost the same performance in the saturation regions as the S-SDR beamforming scheme with much lower complexity.

\begin{figure}
\centering
\subfigure[$\text{RMSE}_1$ vs. $d_{\text{S2S}}$ with different $\rho$ ($\tau_{1}=20,~T=40$).]{\includegraphics[width=0.47\linewidth]{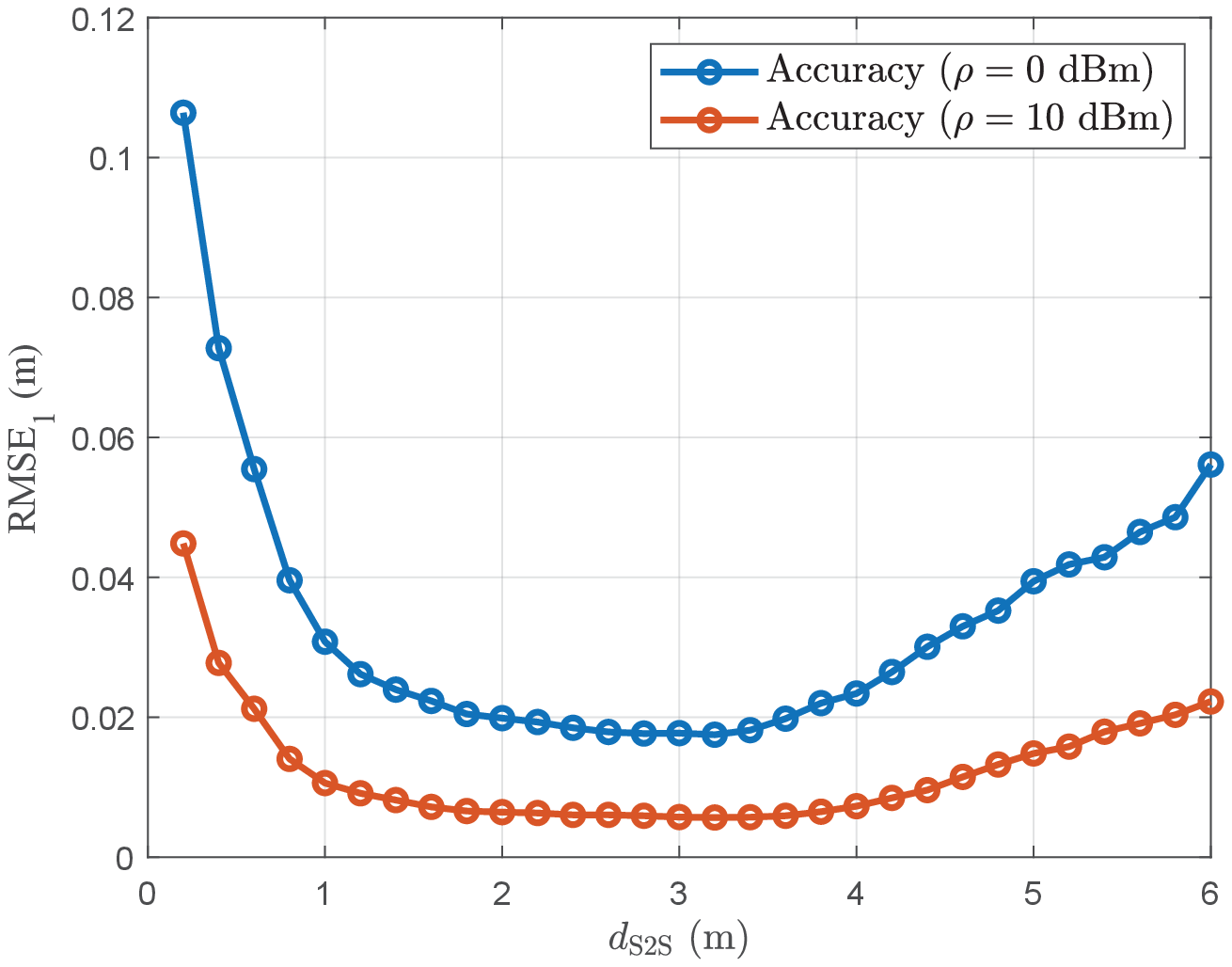}} 
\subfigure[$\bar{R}_2$ vs. $\text{RMSE}_2$ with different $d_{\text{S2S}}$ ($\tau_{1}=20,~T=40,$ $\rho$= 0 dBm).]{\includegraphics[width=0.47\linewidth]{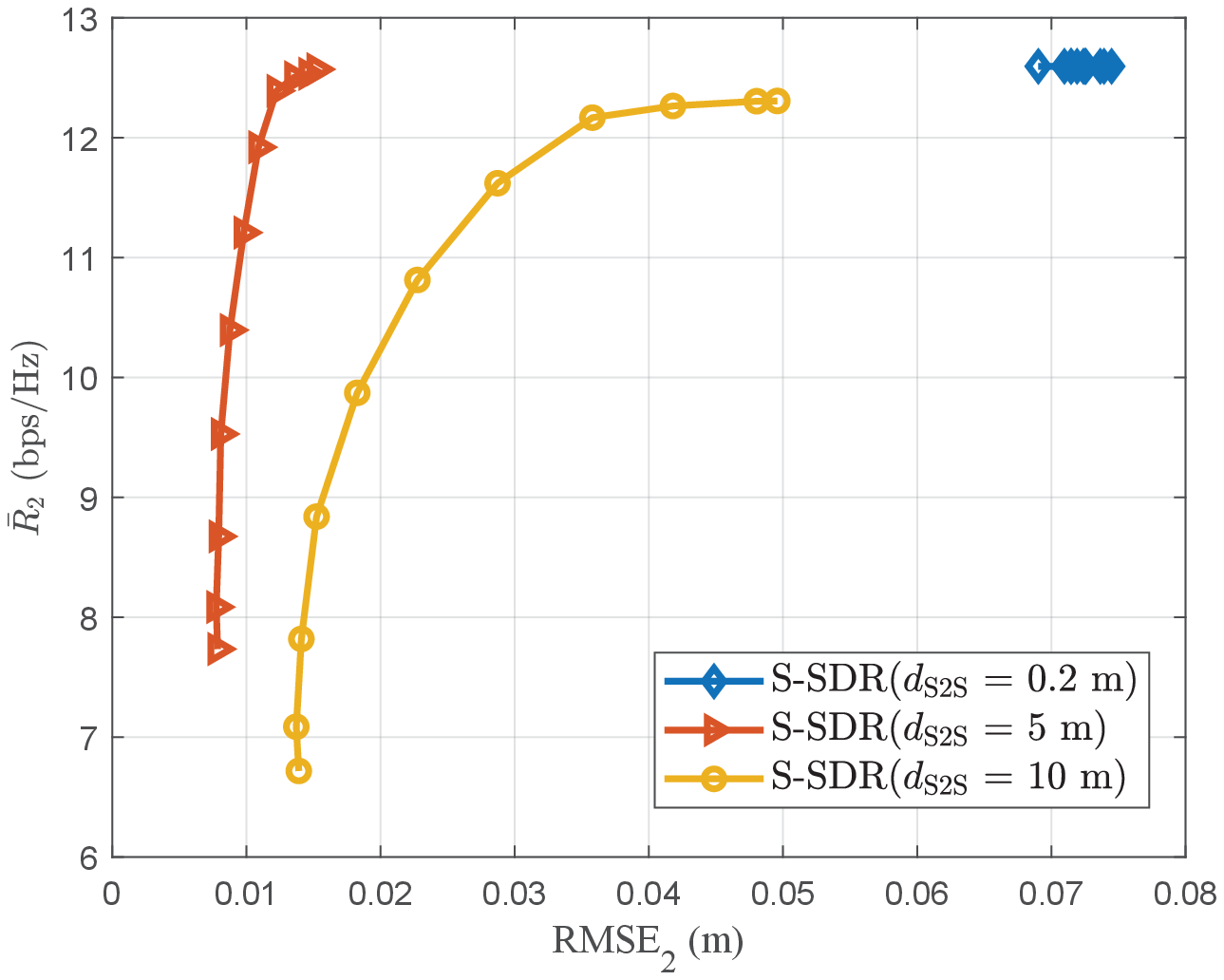} }
\caption{Impact of the positions of the sensing sub-surfaces.}
\label{Response-D-1}
\end{figure}

\textcolor{black}{In Fig. \ref{Response-D-1}, we investigate the impact of the positions of the sensing sub-surfaces. Fig. \ref{Response-D-1}(a) demonstrates the effects on the sensing accuracy in phase 1, where the reflecting sub-surface and the two sensing sub-surfaces are located on the same plane (like on the wall in practice) and have the same height, and two sensing sub-surfaces are symmetric with respect to the reflecting sub-surface. Specifically, Fig.12(a) illustrates the effect of the distance between the two sensing sub-surfaces, i.e., $d_{\text{S2S}}$, on the positioning accuracy in phase 1. We can see that there exists an optimal distance between the two sensing sub-surfaces, which achieves the best sensing performance. The optimal distance should be neither too large nor too small. When the distance becomes too small, the error of location sensing increases dramatically due to the characteristic of AoA-based localization. When the distance gets too large, the error also increases due to the decrease of the received SNRs at the two sensing sub-surfaces. Fig. 12(b) shows the JC\&S performance in phase 2 with different configurations of $d_{\text{S2S}}$. The configuration $d_{\text{S2S}}=5$ m achieves the best JC\&S  performance among the three configurations. A distance that is too large or too small will degrade the JC\&S performance. Therefore, the positions of the sensing sub-surfaces should be properly designed, taking into account both the AoA estimation error due to a small $d_{\text{S2S}}$ and the decreased received SNRs due to a large $d_{\text{S2S}}$.}

\subsection{How Does the Sensing Accuracy Affect the Communication Performance?}\label{simulation_subsection_C}
In this subsection, we focus on answering how the sensing accuracy in phase 1 affects the communication performance in phase 2.



\begin{figure}[htbp]
\centering
\subfigure[$\bar{R}_2$ vs. $M_{\text{s}}$ with different $\rho$ $(\varrho=0,~d_{\text{U2R},1}=5$ m).]
{\includegraphics[width=0.47\linewidth]{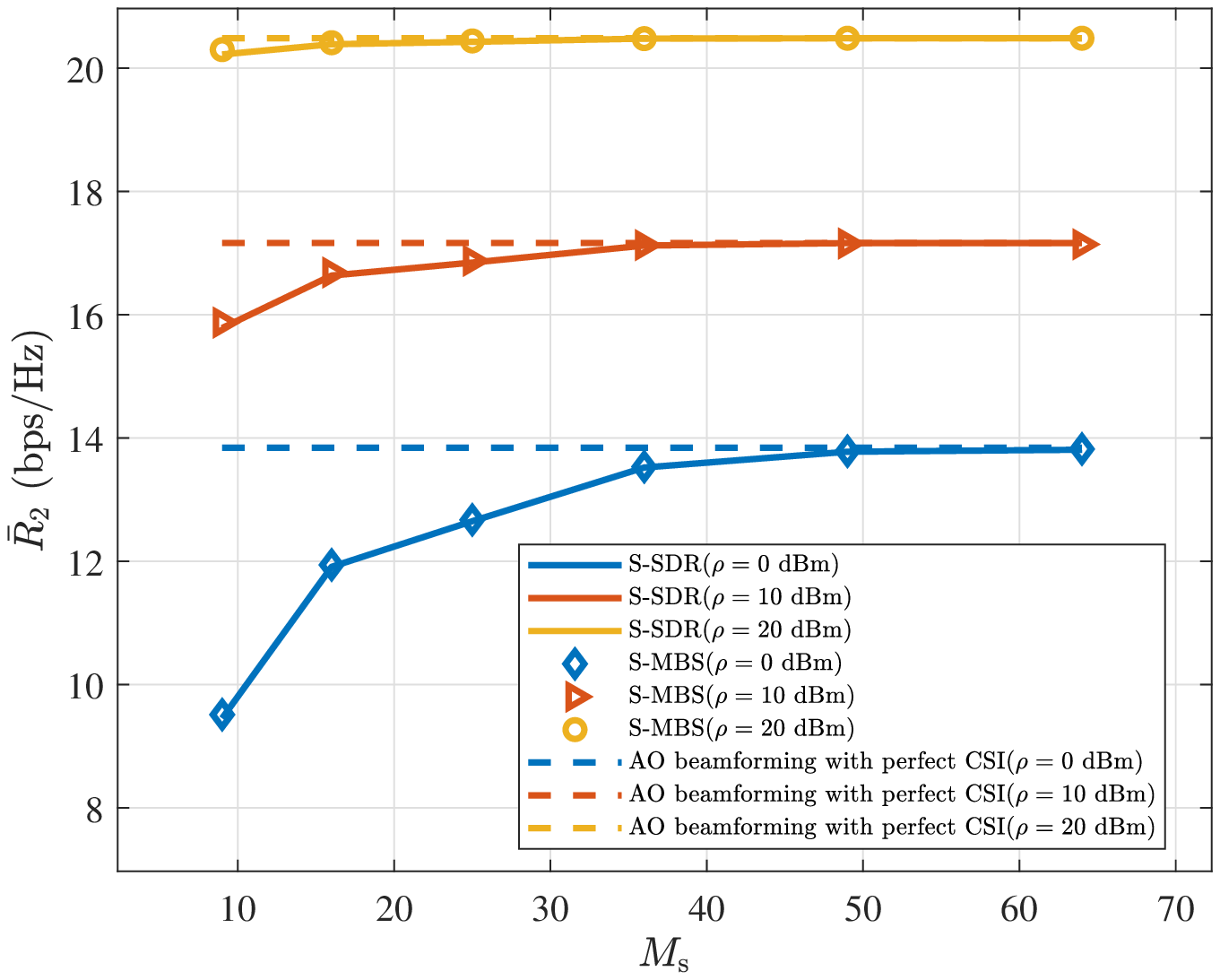} } 
\subfigure[$\bar{R}_2$ vs. $\tau_{1}/T$ with different $M_{\text{s}}$ $(\varrho=0,~\rho=10$ dBm, $T=50$, $d_{\text{U2R},1}=10$ m).]{\includegraphics[width=0.47\linewidth]{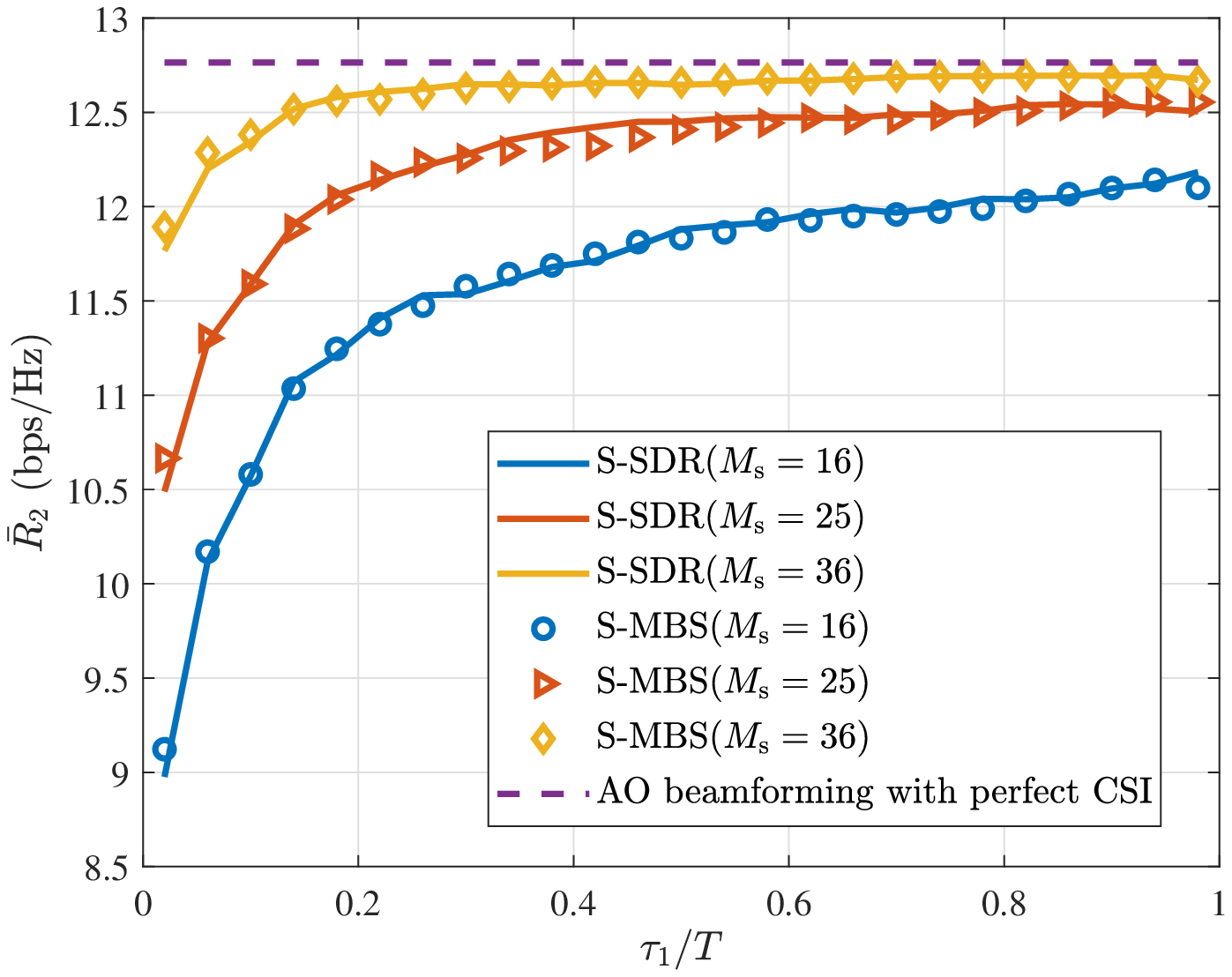} }
\caption{Impact of the sensing accuracy on the communication performance.}
\label{sim_C_2}
\end{figure} 
 
Fig. \ref{sim_C_2} shows the impact of the sensing accuracy in phase 1 on the average communication rate in phase 2 where the trade-off factor $\varrho$ is set to be 0. Fig. \ref{sim_C_2}(a) illustrates that the communication rate of the proposed two beamforming algorithms gradually gets close to that of the AO beamforming algorithm with perfect CSI, when $M_{\text{s}}$ increases. Additionally, as the transmit power or the number of sensing elements increases, the gap of the communication performance between the proposed sensing-based beamforming and the AO beamforming with perfect CSI gradually vanishes, due to the increased positioning accuracy in phase 1. Fig. \ref{sim_C_2}(b) shows that the increase of $\tau_{1}/T$ is able to improve the communication performance of the proposed sensing-based beamforming schemes in phase 2. However, when the number of sensing elements $M_{\text{s}}$ is small, there still exists a communication performance gap between the proposed beamforming and the AO beamforming even with a large number of time slots allocated to phase 1.

\begin{figure}
  \centering  
    \includegraphics[width=0.5\linewidth]{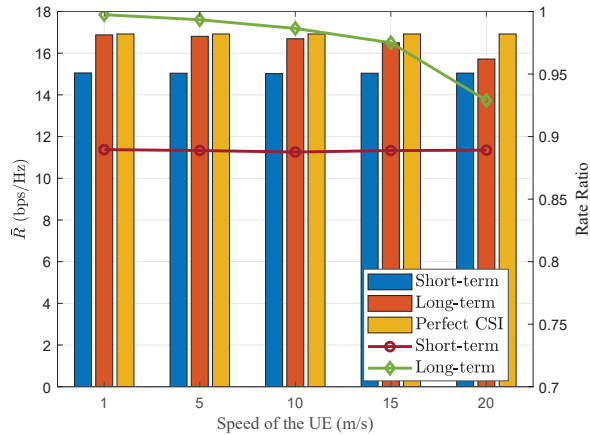}
  \caption{\textcolor{black}{Impact of the UE's mobility on proposed sensing-based beamforming schemes.}}
  \label{Response-E-2}  
\end{figure}  

\textcolor{black}{Fig. \ref{Response-E-2} shows the performance of the proposed sensing-based beamforming schemes when taking into account UE's mobility. In the simulation, we compare three beamforming schemes including the short-term beamforming, long-term beamforming, and beamforming with perfect CSI \cite{wu2019intelligent}. For the short-term scheme, the beamforming is designed according to the transmission protocol in Fig. \ref{transmission protocol}. For the long-term scheme, we use the extended transmission protocol explained in Remark 2. Simulation results are given in the figure with two y-axes, where the left y-axis expresses the average communication rate $\bar{R}$ over multiple coherence blocks, and the right y-axis expresses the rate ratio of the two beamforming schemes, i.e., short-term/long-term beamforming and AO beamforming with perfect CSI. By exploiting the sensing information obtained in the previous block for beamforming in the current block, the proposed long-term scheme performs better than the short-term beamforming scheme. Although $\bar{R}$ of the proposed long-term scheme degrades as the speed of UE increases, it only drops about 7\% when the speed of the UE increases from 1 m/s to 20 m/s.} 

\section{Conclusion}
{In this paper, we constructed an RIS-aided MIMO ISAC framework, where communication and location sensing tasks can be carried out concurrently on the same time-frequency resources. We first proposed a two-phase ISAC transmission protocol. In the first ISAC phase, the ECSI-based active beamforming scheme was used to maximize communication performance, while realizing the coarse-grained location sensing. In the second ISAC phase, by using the coarse-grained sensing information obtained in phase 1, two sensing-based beamforming schemes were proposed to achieve higher-rate communication and fine-grained location sensing simultaneously, while avoiding the high-overhead cascaded channel estimation. Numerical results showed that, the proposed RIS-aided ISAC system can not only achieve almost the same communication performance as the RIS-aided communication system with perfect CSI, but also provide additional high-accuracy location sensing service. Moreover, we has demonstrated the benefit of using sensing information for beamforming design in flexibly balancing and concurrently boosting the communication and sensing performances. In addition, investigation on the communication-sensing trade-off has shown the existence of the balance region and the sensing/communication saturation region, and the ISAC system is supposed to operate in the balance region such that both high-rate communication and high-accuracy sensing can be achieved. Also, improving the sensing accuracy in phase 1 is able to ameliorate the performance of communication in phase 2 when the transmission power is low, which implies that communication can benefit from sensing.}

\ifCLASSOPTIONcaptionsoff
  \newpage
\fi

{\small
\bibliographystyle{IEEEtran}
\bibliography{IEEE_TCOM}{}
}

\end{document}